\documentclass[aps,prfluids,onecolumn,superscriptaddress,amsmath,amssymb]{revtex4-2}

\usepackage{graphicx}
\usepackage{booktabs}
\usepackage{txfonts}
\usepackage[hypertexnames=false]{hyperref}

\newcommand{\dd}{\mathrm{d}}
\newcommand{\ii}{\mathrm{i}}
\newcommand{\ee}{\mathrm{e}}
\newcommand{\EE}{\mathrm{E}}
\newcommand{\Real}{\operatorname{Re}}
\newcommand{\Imag}{\operatorname{Im}}
\newcommand{\zb}{\overline{z}}
\newcommand{\zob}{\overline{z_0}}

\begin{document}

\title{Dynamics of a two-dimensional microswimmer near a slippery boundary}

\author{Takahiro Kanazawa}
\email{kanazawa@uchicago.edu}
\affiliation{Department of Physics, The University of Chicago, 5720 South Ellis Avenue, Chicago, Illinois 60637, USA}
\affiliation{Department of Physics, The University of Tokyo, 7-3-1 Hongo, Bunkyo-ku, Tokyo 113-0033, Japan}
\affiliation{Nonequilibrium Physics of Living Matter Laboratory, RIKEN Pioneering Research Institute, 2-1 Hirosawa, Wako 351-0198, Japan}
\author{Darren Crowdy}
\email{d.crowdy@imperial.ac.uk}
\affiliation{Department of Mathematics, Imperial College London, 180 Queen's Gate, London SW7 2AZ, UK}

\begin{abstract}
Microswimmers near walls and interfaces can switch between qualitatively different modes of motion depending on the slip properties of the neighbouring boundaries. To study this theoretically, a two-dimensional point-singularity model of a microswimmer, which does not self-propel in unbounded fluid, is placed near a flat viscosity discontinuity between two fluids with viscosity ratio $\sigma$, and also near a partial-slip wall with Navier-slip length $\lambda$. Using complex-variable methods, the dynamical systems governing the swimmer motion are derived explicitly in both scenarios. Near the viscosity discontinuity, below a universal critical viscosity ratio $\sigma_c=1/2$, the swimmer is trapped at a fixed height with its stresslet axis normal to the boundary. When $\sigma=\sigma_c$ it glides parallel to the boundary at a speed selected by its orientation, with a terminal speed determined by its initial state. For $\sigma>\sigma_c$ it oscillates periodically in height and orientation while drifting along the boundary. The dynamics near a partial-slip wall depend on the ratio of the slip length to the length scale of the singularity and are similar in exhibiting an analogous transition at a critical slip length. It is found, however, that the dynamics near a viscosity discontinuity for $\sigma$ above $\sigma_c$ are richer than those near the partial-slip boundary, including the existence of a hydrodynamic bound state with zero net displacement over one period. Although the study is limited to two dimensions, these results show how a single boundary parameter organises qualitatively distinct modes of locomotion, in close analogy with slip-induced transitions reported for three-dimensional swimmers.
\end{abstract}

\maketitle

\section{Introduction}\label{sec:intro}

Swimming microorganisms and synthetic microswimmers often move near solid walls and fluid interfaces, where hydrodynamic interactions with the boundary lead to accumulation, reorientation, sliding, hovering and circular trajectories~\cite{lauga2006swimming,berke2008hydrodynamic,lauga2009hydrodynamics,di2011swimming,spagnolie2012hydrodynamics,elgeti2016microswimmers,bechinger2016active}. Such boundary-induced dynamics influence processes ranging from biofilm formation to the design of microrobots for targeted transport, and a large body of work has examined how the swimmer's gait and shape control its behaviour near a no-slip wall~\cite{lauga2006swimming,berke2008hydrodynamic,ishimoto2013squirmer,li2014hydrodynamic,spagnolie2012hydrodynamics,uspal2015self,htet2024hydrodynamic}.

Real boundaries are often neither perfectly rigid nor free of shear. Liquid--liquid interfaces separate fluids of different viscosity, and the effective boundary condition at a solid surface can involve hydrodynamic slip, for example on hydrophobic or polymer-depleted substrates~\cite{lauga2005brownian,lemelle2013curvature,hu2015physical,ketzetzi2020slip}. For passive bodies near a liquid--liquid interface, Fulford and Blake~\cite{fulford1983motion} computed the drag and interface-induced rotation of a slender body, and found that for a narrow band of viscosity ratios the sense of rotation changes with distance from the interface, suggesting a possible capture region. Far-field singularity models show that a clean fluid interface, a slipping wall and a surfactant-laden interface all attract a pusher swimmer and align it parallel to the surface~\cite{lopez2014dynamics}. Closer to the boundary the picture is richer. Simulations of squirmers near liquid--liquid interfaces reveal bouncing, sliding and penetrating modes, together with a hovering state that appears only when the two viscosities differ~\cite{feng2022dynamics,feng2023dynamics}.

For partial-slip walls, exact solutions for spherical squirmers show that swimmers which escape a no-slip wall become trapped near it, in periodic or damped oscillations or in steady sliding, once the slip length exceeds a critical value~\cite{poddar2020near}. Slip-induced transitions have since been reported for catalytic micromotors, which switch sharply from escape to wall collision or stable sliding~\cite{ghosh2026slippery}, for squirmers in shear and Poiseuille flows near slippery walls~\cite{ghosh2023slippery,ghosh2024harnessing}, and in lattice Boltzmann simulations of elongated swimmers, for which wall-bound sliding or oscillating states appear beyond a critical slip length~\cite{guan2024effect}. Slip can likewise change the type of motion of passive particles: thin nanoplatelets with hydrodynamic slip stop rotating in shear flow and align at a small tilt angle~\cite{kamal2020hydrodynamic,gravelle2021violations}, and an exact two-dimensional analysis shows that a rigid rod of elliptical cross-section with Navier slip, which would rotate periodically in Jeffery orbits without slip, attains an equilibrium tilt once the slip length exceeds a critical value set by its aspect ratio~\cite{crowdy2022equilibrium}.

These transitions are typically identified by integrating trajectories numerically for many parameter values. It is then difficult to see which features of the flow are responsible for the transition, and whether it reflects a generic bifurcation or a property of a particular numerical model. A complementary approach is to seek minimal models in which the swimmer dynamics can be written down in closed form, so that transitions can be traced to explicit changes in the structure of a low-dimensional dynamical system.

Two-dimensional models are well suited to this purpose. Complex-variable methods for Stokes flow give exact solutions in geometries where three-dimensional calculations require heavy numerics. Crowdy and Or~\cite{crowdy2010two} introduced a point-singularity model of a circular treadmilling swimmer near a no-slip wall and found periodic, wavy trajectories along the wall; the same qualitative dynamics were later obtained from an exact solution for the finite-sized swimmer~\cite{crowdy2011treadmilling} and extended to background shear~\cite{ishimoto2017dynamics}. The approach has been used for swimmers beneath a free surface~\cite{crowdy2011two}, and near a gap in a wall and in a corner, where hydrodynamic bound states appear~\cite{crowdy2011hydrodynamic,davis2012stresslet}. Systematic corrections to the point-singularity model, correct to third order in the swimmer radius, have also been derived~\cite{davis2013matched}. Qualitatively similar periodic near-wall motions have been found for swimmers made of linked spheres~\cite{or2009dynamics,or2011dynamics}, for a spheroidal squirmer~\cite{ishimoto2013squirmer} and in experiments with small robotic swimmers~\cite{zhang2010experimental}. A direct comparison of disk and spherical squirmers near a wall also finds a similar bifurcation structure in two and three dimensions, although the stability of some fixed points differs~\cite{ishimoto2017guidance}. These results support the use of idealised two-dimensional models as a source of physical insight.

In this paper we use this framework to study how two kinds of complex boundary, a flat interface between fluids of different viscosity and a partial-slip wall, change the dynamics of the treadmilling swimmer of \cite{crowdy2010two}. This swimmer is active but not self-propelled; in the terminology of active suspensions it is a ``shaker'' rather than a ``mover''~\cite{hatwalne2004rheology}, and related models have been called stirrers~\cite{yazdi2015swimming}. We consider this swimmer for simplicity: it is the simplest singularity model that interacts nontrivially with a boundary, and since it does not move in unbounded fluid, all of its motion arises from the boundary. Self-propelled swimmers can be treated within the same framework. We derive closed-form ordinary differential equations for the swimmer's position and orientation for arbitrary viscosity ratio $\sigma$, and in terms of exponential integrals for arbitrary slip length $\lambda$.

Our main findings are as follows. (i) For both boundaries, every qualitative change in the dynamics corresponds to the crossing of two of three characteristic heights, at which the velocity parallel to the boundary, the velocity normal to it, or the rotation rate of the swimmer vanishes. None of these changes occurs for the stresslet alone: they are produced by the source quadrupole. (ii) Below a critical viscosity ratio $\sigma_c=1/2$, which is independent of the length scale set by the relative strength of the quadrupole and the stresslet, the swimmer is trapped at a fixed height with its axis normal to the interface and comes to rest. (iii) At $\sigma=\sigma_c$ the swimmer glides parallel to the interface at a speed selected by its orientation alone, and its terminal speed can be predicted exactly from its initial position and orientation. (iv) For $\sigma>\sigma_c$ the motion consists of periodic oscillations in height and orientation, superposed on a mean drift along the interface; the drift of small orbits reverses direction at $\sigma=1$, so hydrodynamic bound states with zero mean drift exist only for $1/2<\sigma<1$, and trajectories that loop backwards exist only for $1/2<\sigma<(1+\sqrt3)/2$. (v) Near a partial-slip wall the same primary transition occurs at a critical slip length proportional to the length scale of the singularity, giving an explicit two-dimensional counterpart of the slip-induced transitions reported for three-dimensional swimmers. The secondary crossings, however, occur in the reverse order: the slip wall has no looping trajectories or bound states, but has a range of slip lengths in which no trapped swimmer reverses, and at the transition the swimmer glides in the opposite direction.

The paper is organised as follows. Section~\ref{sec:model} introduces the swimmer model, the boundary conditions and the resulting equations of motion; derivations are given in the appendices. Section~\ref{sec:interface} analyses the dynamics near a fluid--fluid interface. Section~\ref{sec:slip} treats the partial-slip wall and compares the two boundaries. Section~\ref{sec:discussion} discusses the physical interpretation, the limitations of the model and its relation to three-dimensional studies.

\section{Point-singularity swimmer model}
\label{sec:model}

The aim of this study is to write down, and study, the dynamical system governing the dynamics of a swimmer, modelled as a point singularity of the Stokes equations, located in the upper half plane $y>0$ near a flat slippery boundary at $y=0$. The point singularity to be studied is a superposition of a stresslet of (non-dimensional) strength $\mu \in \mathbb{C}$ and a source quadrupole of strength $\beta \in \mathbb{C}$. These two complex singularity strengths are in correspondence with two real parameters, called $\theta$ and $\epsilon$, according to
\begin{equation}
\mu=\ee^{2\ii\theta}, \qquad \beta=2\epsilon^2 \mu.
\end{equation}
It is explained in Appendix~\ref{app:point_swimmer} why this particular choice of singularity is made: it corresponds to a singularity description, in free space, of a particular type of circular treadmilling swimmer first considered by Crowdy and Or~\cite{crowdy2010two} who studied its dynamics in the presence of a no-slip wall. The present study can be viewed as a generalisation of that work to studying the same swimmer near a slippery boundary of the two types to be described next. The angle $\theta$ is a proxy for the orientation of the swimmer and the parameter $\epsilon$, which measures the quadrupole strength relative to the stresslet strength, is a proxy for the swimmer size, although it is to be remembered that the point singularity has no spatial extent.

It should be noted that it is more common in the literature to use the symbol $\beta$ to characterise the relative strength of the stresslet to a source dipole contribution (rather than the source quadrupole used here), in which case, the terms ``pusher'' and ``puller'' are often used. This is different to what is done in this paper, although the methodology used here can be extended to this case and is the topic of ongoing work.

\subsection{Two types of slippery boundary}\label{sec:bc}

\begin{figure}[t]
\centering
\includegraphics[width=0.75\textwidth]{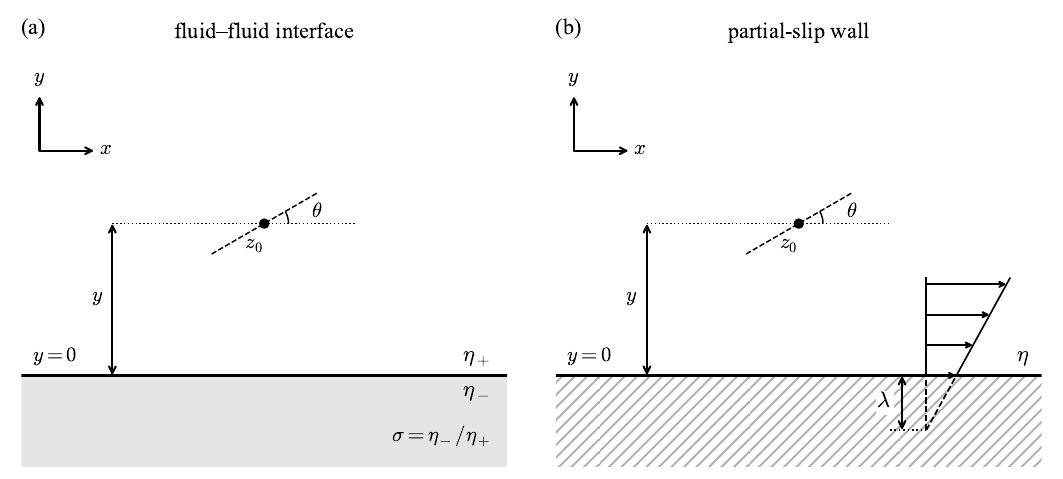}
\caption{A stresslet-quadrupole singularity near two types of slippery boundary. (a)~The boundary is an impermeable interface between the upper fluid, of viscosity $\eta_+$, in which the point singularity is located and a second fluid of viscosity $\eta_-$, with continuity of velocity and shear stress at $y=0$; $\sigma=\eta_-/\eta_+$. (b)~The boundary is a rigid wall with a Navier-slip boundary condition with slip length $\lambda$.}
\label{fig:schematic}
\end{figure}

\paragraph{Fluid--fluid interface.} In this case, the swimmer is taken to lie in the upper fluid, of viscosity $\eta_+$, with the lower half plane occupied by a second fluid of viscosity $\eta_-$ (Fig.~\ref{fig:schematic}a). The viscosity ratio is
\begin{equation}
\sigma=\frac{\eta_-}{\eta_+} .
\label{eq:sigma}
\end{equation}
The flat interface at $y=0$ is taken to be impermeable, corresponding to a limit of vanishing capillary number, with continuity of velocity and of shear stress imposed across it. The limits $\sigma\to\infty$ and $\sigma\to0$ correspond to a no-slip wall and a free surface, respectively. Note that $\sigma=1$ is not a homogeneous fluid: it describes a flat, non-deformable interface separating two fluids of equal viscosity, on which a distribution of normal stress is implied, so the interface still affects the flow~\cite{aderogba1978action}. Such an interface could arise, for example, between two fluids of equal viscosity but different density~\cite{fulford1983motion} or, as supposed here, if strong surface tension is active on the interface.

\paragraph{Partial-slip wall.} A second type of slippery boundary to be considered is that of a rigid wall at which the fluid obeys the Navier-slip condition
\begin{equation}
u\big|_{y=0}=\lambda\,\frac{\partial u}{\partial y}\bigg|_{y=0},\qquad v\big|_{y=0}=0,
\label{eq:navier}
\end{equation}
where $u$ and $v$ are the components of the fluid velocity in the $x$ and $y$ directions and $\lambda$ is the slip length (Fig.~\ref{fig:schematic}b). 
This model introduces a new length scale into the problem so, since $\epsilon$ is the proxy for the swimmer radius, the relevant new parameter is the ratio $\lambda/\epsilon$ (Section~\ref{sec:odes_slip}). The limits $\lambda\to0$ and $\lambda\to\infty$ give the no-slip wall and the shear-free (free-surface) boundary.

\subsection{Complex-variable formulation}\label{sec:method}

To solve for the Stokes flow generated by the point singularity near these boundaries, the complex variable formulation of two-dimensional Stokes flow~\cite{flotilla, crowdy2020solving} is employed. An appendix of Crowdy~\cite{flotilla} gives a derivation of the governing equations to be stated next. With $z=x+\ii y$ and an overbar denoting complex conjugation, the biharmonic streamfunction $\psi$ has a representation
\begin{equation}
\psi=\Imag[\zb f(z)+g(z)],
\end{equation}
in terms of two functions, $f(z)$ and $g(z)$, which are analytic in the fluid region. It should be mentioned that these analytic functions will depend on time as the singularity moves around and its strength changes, but since the problem is quasi-steady, it is natural to suppress this dependence on time.
In a fluid of viscosity $\eta$, the velocity $(u,v) = (\partial \psi/\partial y, -\partial \psi/\partial x)$, fluid pressure $p$ and 
vorticity $\omega = - \nabla^2 \psi$ are given by
\begin{equation}
u-\ii v=-\overline{f(z)}+\zb f'(z)+g'(z),\qquad \frac{p}{\eta}-\ii\omega=4f'(z).
\label{eq:goursat}
\end{equation}
The challenge is to find $f(z)$ and $g(z)$. A key step is to write 
\begin{equation}
g(z)=-zf(z)+\hat{g}(z)
\end{equation}
in each fluid  and it is easily seen  that $y=0$ is a streamline provided that
\begin{equation}
\Imag[\hat{g}'(z)] = 0 \qquad \text{on}~y=0.
\end{equation}
The relation~\eqref{eq:goursat} can then be used to show that the slip velocity along the boundary is $\Real[h(z)]$ where 
\begin{equation}
h(z) \equiv -2f(z)+\hat{g}'(z).
\label{eq:h}
\end{equation}
For the first case of an interface between two fluids of different viscosity $\eta_\pm$, each fluid has its own analytic functions, denoted $f_\pm(z)$, $g_\pm(z)$, $\hat{g}_\pm(z)$ and $h_\pm(z)$. In the second problem of the partial-slip wall, there is a single fluid and the subscript is dropped. The point is that $h(z)$ and $\hat{g}(z)$ are the functions to be determined, from which $f(z)$ and $g(z)$ follow.

Full derivations of the governing dynamical systems are given in Appendix~\ref{app:interface} for the viscosity discontinuity and Appendix~\ref{app:slip} for the partial-slip wall, but the main ideas are as follows. The stresslet-quadrupole combination representing the swimmer fixes the local behaviour of $h(z)$ in the upper fluid near $z_0$, indeed, the general form is given by
\begin{equation}
h_+(z)=\frac{A}{z-z_0}+\frac{B}{(z-z_0)^2}+\frac{C}{(z-z_0)^3}+(\text{analytic}),
\label{eq:hlocal}
\end{equation}
with
\begin{equation}
A=-2\mu,\qquad B=\mu(\zob-z_0),\qquad C=\beta .
\end{equation}

For the two-fluid interface, continuity of the slip velocity and of the shear stress become 
\begin{equation}
\Real [h_+(z)]=\Real [h_-(z)],\qquad \Imag [h_+(z)]=\sigma\Imag [h_-(z)]
\end{equation}
on $y=0$. A Schwarz-reflection argument (Appendix~\ref{app:interface}) gives closed-form expressions for the required
analytic functions as follows:
\begin{equation}
\begin{split}
\hat{g}_+'(z)&=\frac{B}{(z-z_0)^2}+\frac{C}{(z-z_0)^3}+\frac{\overline{B}}{(z-\zob)^2}+\frac{\overline{C}}{(z-\zob)^3}, \\
h_+(z)&=\frac{A}{z-z_0}+\frac{B}{(z-z_0)^2}+\frac{C}{(z-z_0)^3}
+\frac{1-\sigma}{1+\sigma}\left[\frac{\overline{A}}{z-\zob}+\frac{\overline{B}}{(z-\zob)^2}+\frac{\overline{C}}{(z-\zob)^3}\right].
\end{split}
\label{eq:hsol}
\end{equation}

For the partial-slip wall, the Navier-slip condition~\eqref{eq:navier} can be shown to be equivalent to a first-order linear differential equation for $h(z)$ on the real axis, which can be solved using an integrating factor. The solution for $h(z)$  is then expressible in terms of exponential integrals $\EE_n$ (Appendix~\ref{app:slip}). 

Because the swimmer is force- and torque-free, it moves with the regular part of the image-generated flow at its location~\cite{crowdy2010two}. Near $z_0$ we write
\begin{equation}
\begin{aligned}
u-\ii v&=(\text{singular})+U-\ii V+O(z-z_0),\\
\frac{p}{\eta_+}-\ii\omega&=(\text{singular})+\frac{p_0}{\eta_+}-\ii\omega_0+O(z-z_0).
\end{aligned}
\label{eq:regular}
\end{equation}
The real constant $p_0$ is the regular part of the pressure, which does not enter the swimmer dynamics. The equations of motion of the swimmer are then
\begin{equation}
\frac{\dd x}{\dd t}=U,\qquad \frac{\dd y}{\dd t}=V,\qquad \frac{\dd\theta}{\dd t}=\frac{\omega_0}{2}.
\label{eq:faxen}
\end{equation}
The angular velocity of the swimmer axis is half the regular part of the vorticity at the swimmer location.

\subsection{Equations of motion: fluid--fluid interface}\label{sec:odes_interface}

Evaluating the regular parts~\eqref{eq:regular} using \eqref{eq:hsol}, the governing dynamical system is found to be
\begin{subequations}\label{eq:ode}
\begin{align}
\frac{\dd x}{\dd t}&=-\frac{\sigma}{1+\sigma}\frac{\sin{2\theta}}{y}\left(1-\frac{y_x^2}{y^2}\right),\label{eq:xdot}\\
\frac{\dd y}{\dd t}&=\frac{\cos{2\theta}}{y}\left(1-\frac{y_y^2}{y^2}\right),\label{eq:ydot}\\
\frac{\dd\theta}{\dd t}&=\frac{\sin{2\theta}}{2y^2}\left(1-\frac{y_\theta^2}{y^2}\right),\label{eq:thdot}
\end{align}
\end{subequations}
where the positive heights $y_x$, $y_y$ and $y_\theta$ are given by
\begin{equation}
y_x^2=\frac{1+2\sigma}{4\sigma}\,\epsilon^2,\qquad
y_y^2=\frac{1+4\sigma}{4(1+\sigma)}\,\epsilon^2,\qquad
y_\theta^2=\frac{3\sigma}{2(1+\sigma)}\,\epsilon^2 .
\label{eq:heights}
\end{equation}
Thus $y_x$, $y_y$ and $y_\theta$ are the heights at which the velocity parallel to the boundary, the velocity normal to it and the rotation rate vanish for any orientation. As $\sigma\to0$, $y_x\to\infty$ but the product in \eqref{eq:xdot} remains finite, giving $\dot{x}=\epsilon^2\sin2\theta/(4y^3)$ for a free surface, where a dot denotes $\dd/\dd t$.

Several features of \eqref{eq:ode} are worth noting at the outset. Because the boundary is flat, the problem is invariant under translations along it, so the right-hand sides of \eqref{eq:ode} are independent of $x$. The $(y,\theta)$ subsystem~\eqref{eq:ydot}--\eqref{eq:thdot} can therefore be studied on its own, and the position along the boundary follows by direct integration of \eqref{eq:xdot} along a solution $(y(t),\theta(t))$:
\begin{equation}
x(t)=x(0)+\int_0^t \dot x\bigl(y(t'),\theta(t')\bigr)\,\dd t'.
\end{equation}
The dynamics are $\pi$-periodic in $\theta$ and invariant under $(x,\theta)\mapsto(-x,-\theta)$ by reflection symmetry about the normal to the boundary. As $\sigma\to\infty$, \eqref{eq:ode} reduces to the no-slip-wall equations of Crowdy and Or~\cite{crowdy2010two}. 

The parameter $\epsilon$ can be scaled out of \eqref{eq:ode}: if $(X(t),Y(t),\Theta(t))$ is a solution for $\epsilon=1$, then $(\epsilon X(t/\epsilon^2),\,\epsilon Y(t/\epsilon^2),\,\Theta(t/\epsilon^2))$ is a solution for a general value of $\epsilon$. Changing $\epsilon$ 
only stretches lengths by a factor $\epsilon$ and times by $\epsilon^2$. Every characteristic height therefore scales with $\epsilon$, and every critical viscosity ratio found below is independent of $\epsilon$. 

\subsection{Equations of motion: partial-slip wall}\label{sec:odes_slip}

For the partial-slip wall it is generally natural to measure lengths in units of the slip length, $\tilde{y}=y/\lambda$ and $\tilde{\epsilon}=\epsilon/\lambda$. The equations of motion take the form
\begin{equation}
\frac{\dd x}{\dd t}=-\frac{\sin{2\theta}}{2\lambda}\,\mathcal{F}_x(\tilde{y};\tilde{\epsilon}),\qquad
\frac{\dd y}{\dd t}=\frac{\cos{2\theta}}{2\lambda}\,\mathcal{F}_y(\tilde{y};\tilde{\epsilon}),\qquad
\frac{\dd\theta}{\dd t}=\frac{\sin{2\theta}}{2\lambda^2}\,\mathcal{F}_\theta(\tilde{y};\tilde{\epsilon}),
\label{eq:ode_slip}
\end{equation}
with
\begin{align}
\mathcal{F}_x(\tilde{y};\tilde{\epsilon})&=\ee^{\tilde y}\EE_1(\tilde{y})\left[\tilde{y}^2+\left(4+\frac{\tilde{\epsilon}^2}{4}\right)\tilde{y}+4+\frac{\tilde{\epsilon}^2}{2}\right] -\tilde{y}-3-\frac{\tilde{\epsilon}^2}{4}-\frac{\tilde{\epsilon}^2}{4\tilde{y}}-\frac{\tilde{\epsilon}^2}{2\tilde{y}^3},\label{eq:Fxslip}\\
\mathcal{F}_y(\tilde{y};\tilde{\epsilon})&=\tilde{y}\,\ee^{\tilde{y}}\EE_1(\tilde{y})\left(\tilde{y}+2+\frac{\tilde{\epsilon}^2}{4}\right) -\tilde{y}-1-\frac{\tilde{\epsilon}^2}{4}+\left(2+\frac{\tilde{\epsilon}^2}{4}\right)\frac{1}{\tilde{y}}-\frac{\tilde{\epsilon}^2}{2\tilde{y}^2}-\frac{\tilde{\epsilon}^2}{2\tilde{y}^3},\label{eq:Fyslip}\\
\mathcal{F}_\theta(\tilde{y};\tilde{\epsilon})&=\ee^{\tilde y}\EE_1(\tilde{y})\left(\tilde{y}+2+\frac{\tilde{\epsilon}^2}{4}\right) -1-\left(1+\frac{\tilde\epsilon^2}{4}\right)\frac{1}{\tilde{y}}+\left(1+\frac{\tilde{\epsilon}^2}{4}\right)\frac{1}{\tilde{y}^2}-\frac{\tilde{\epsilon}^2}{2\tilde{y}^3},\label{eq:Fthetaslip}
\end{align}
where $\EE_1(\tilde{y})=\int_1^\infty \ee^{-\tilde{y} s}s^{-1}\,\dd s$ is the exponential integral. The $\theta$-dependence is identical to that of \eqref{eq:ode}, through $\sin{2\theta}$ and $\cos{2\theta}$ only. As $\lambda\to0$ and $\lambda\to\infty$, \eqref{eq:ode_slip} reduces to \eqref{eq:ode} with $\sigma\to\infty$ and $\sigma\to0$, respectively (Appendix~\ref{app:asym}).

The scaling argument of Section~\ref{sec:odes_interface} still applies if the slip length is rescaled with $\epsilon$: if $(X(t),Y(t),\Theta(t))$ is a solution of~\eqref{eq:ode_slip} for $\epsilon=1$ and slip length $\Lambda$, then $(\epsilon X(t/\epsilon^2),\,\epsilon Y(t/\epsilon^2),\,\Theta(t/\epsilon^2))$ is a solution for general $\epsilon$ and slip length $\lambda=\epsilon\Lambda$. The dynamics therefore depend on the single parameter $\lambda/\epsilon$, equivalently $\tilde{\epsilon}=\epsilon/\lambda$, and any critical value corresponds to a critical slip length proportional to $\epsilon$. Although~\eqref{eq:ode_slip} is written in units of $\lambda$, which is natural for the exponential integrals, we report all slip-wall results in units of $\epsilon$, as for the interface, with $\lambda/\epsilon$ as the control parameter.

\section{Swimmer near a fluid--fluid interface}\label{sec:interface}

Before considering the general case, it is instructive to first remove the source quadrupole ($\epsilon\to0$), leaving a pure stresslet. Since the quadrupole contributions in~\eqref{eq:ode} are smaller than the stresslet contributions by a factor $\epsilon^2/y^2$, this also describes the dynamics of a swimmer far from the boundary, $y\gg\epsilon$. Then \eqref{eq:ode} reduces to
\begin{equation}
\dot{x}=-\frac{\sigma}{1+\sigma}\frac{\sin{2\theta}}{y},\qquad \dot{y}=\frac{\cos{2\theta}}{y},\qquad \dot{\theta}=\frac{\sin{2\theta}}{2y^2}.
\label{eq:ode_stresslet}
\end{equation}
The $(y,\theta)$ dynamics are independent of $\sigma$: the interface only rescales the drift parallel to the boundary by the factor $\sigma/(1+\sigma)$, equivalent to a stretch of the $x$ axis. The swimmer reorients towards $\theta=\pm\pi/2$ (its axis normal to the boundary) and reaches the boundary in finite time unless $\theta=0$ (mod $\pi$) exactly. For a free surface ($\sigma=0$) the stresslet does not move parallel to the boundary at all (small capillary number effects on this problem were considered in \cite{crowdy2011two}). Thus, at the level of the leading far-field singularity, a viscosity contrast cannot change whether the swimmer is attracted, nor how it reorients. All the $\sigma$-dependent phenomena described below arise from the source quadrupole. This complements far-field models, in which different interface types induce qualitatively similar attraction and alignment~\cite{lopez2014dynamics}.

\subsection{Three characteristic heights}\label{sec:heights}

\begin{figure}[t]
\centering
\includegraphics[width=0.6\textwidth]{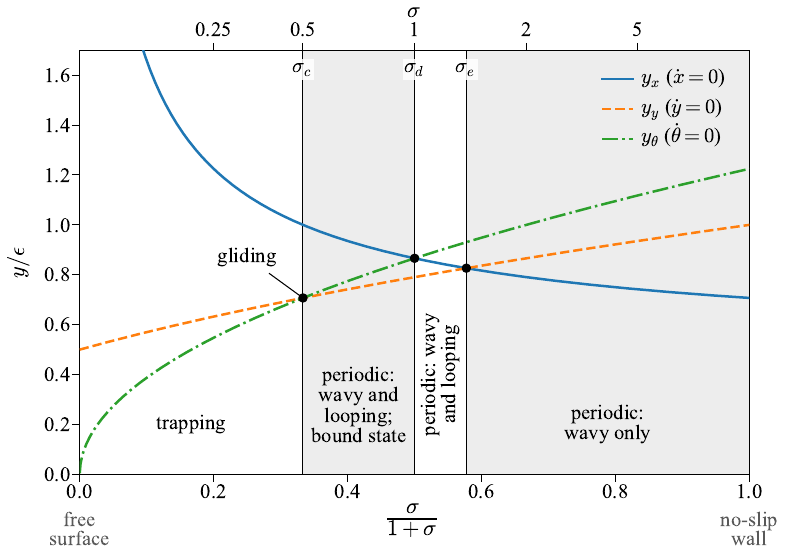}
\caption{Regime diagram for the swimmer near a fluid--fluid interface. The heights $y_x$, $y_y$ and $y_\theta$ of \eqref{eq:heights}, in units of $\epsilon$, are plotted against $\sigma/(1+\sigma)$, which runs from the free surface ($0$) to the no-slip wall ($1$); values of $\sigma$ are shown on the upper axis. The crossings $y_y=y_\theta$, $y_x=y_\theta$ and $y_x=y_y$ (dots) define $\sigma_c=1/2$, $\sigma_d=1$ and $\sigma_e=(1+\sqrt3)/2$. Labels indicate the long-time behaviour in each regime; at $\sigma=\sigma_d$ all orbits loop and the centre is stationary, and for $\sigma\ge\sigma_e$ all orbits are wavy.}
\label{fig:regime}
\end{figure}

To understand the dynamics with the quadrupole restored, it is useful to observe that the factors in \eqref{eq:ode} change sign at the heights~\eqref{eq:heights}. Each qualitative change in the dynamics corresponds to two of these heights crossing (Fig.~\ref{fig:regime}). Equating them pairwise gives three critical viscosity ratios,
\begin{equation}
y_y=y_\theta:\ \sigma_c=\tfrac12,\qquad
y_x=y_\theta:\ \sigma_d=1,\qquad
y_x=y_y:\ \sigma_e=\tfrac{1+\sqrt3}{2}\approx1.366,
\label{eq:critical}
\end{equation}
and the ordering
\begin{equation}
\begin{array}{ll}
y_\theta<y_y<y_x, & 0\le\sigma<\sigma_c,\\
y_y<y_\theta<y_x, & \sigma_c<\sigma<\sigma_d,\\
y_y<y_x<y_\theta, & \sigma_d<\sigma<\sigma_e,\\
y_x<y_y<y_\theta, & \sigma>\sigma_e .
\end{array}
\label{eq:ordering}
\end{equation}
The crossing at $\sigma_c$ changes the equilibria of the $(y,\theta)$ subsystem; the crossings at $\sigma_d$ and $\sigma_e$ do not, but they change the sign structure of the drift parallel to the boundary, and hence the trajectories in the physical plane $(x,y)$. Figure~\ref{fig:regime} gives a summary of the various regimes that are now discussed separately.

\subsection{First integral and equilibria}\label{sec:equilibria}

A route to understanding the dynamics is to make use of a first integral. Dividing \eqref{eq:thdot} by \eqref{eq:ydot} and integrating shows that the $(y,\theta)$ dynamics possess the first integral
\begin{equation}
H(y,\theta)=\frac{\sin{2\theta}}{y^{6\sigma/(1+4\sigma)}\,\left(y^2-y_y^2\right)^{(1-2\sigma)/[2(1+4\sigma)]}},\qquad y>y_y,
\label{eq:firstintegral}
\end{equation}
for all $\sigma$ (full details are given in Appendix~\ref{app:stability}); below $y_y$ the same expression holds with $y^2-y_y^2$ replaced by $y_y^2-y^2$. Writing $H(y,\theta)=\sin{2\theta}/\mathcal{H}(y)$, with
\begin{equation}
\mathcal{H}(y)=y^{6\sigma/(1+4\sigma)}\left(y^2-y_y^2\right)^{(1-2\sigma)/[2(1+4\sigma)]},
\label{eq:calH}
\end{equation}
trajectories of the $(y,\theta)$ subsystem lie on the curves $\sin2\theta=H_0\mathcal{H}(y)$, $H_0$ constant, which we use below to characterise the trapped and gliding states, the periodic orbits and the mean drift.

\begin{figure}[t]
\centering
\includegraphics[width=\textwidth]{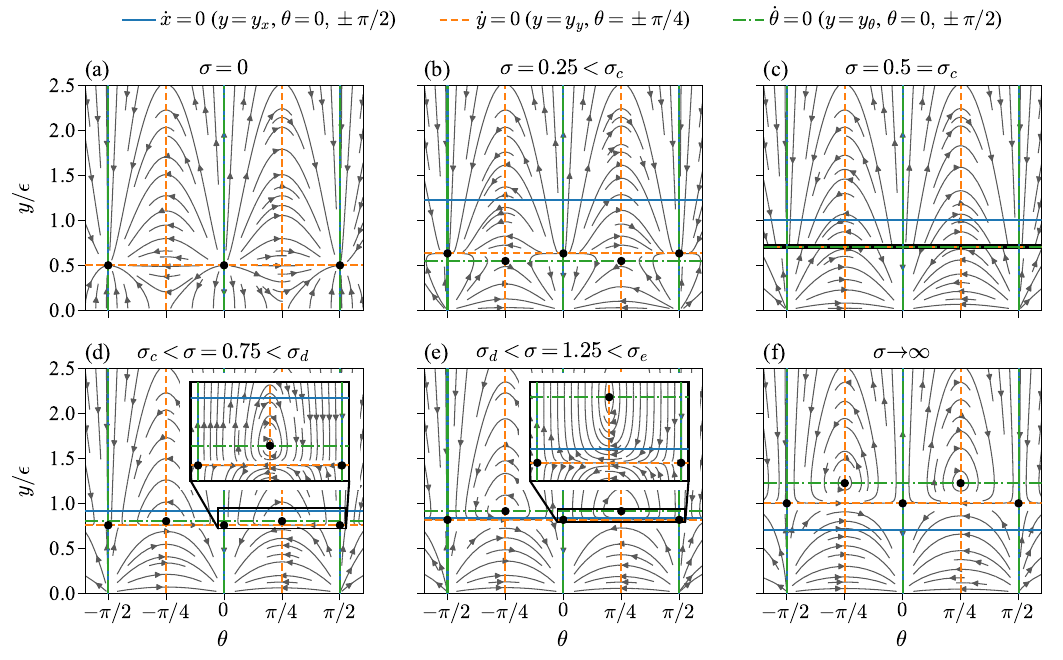}
\caption{Phase portraits of the $(y,\theta)$ subsystem~\eqref{eq:ydot}--\eqref{eq:thdot} for (a)~$\sigma=0$ (free surface), (b)~$0.25$, (c)~$\sigma_c=1/2$, (d)~$0.75$, (e)~$1.25$ and (f)~$\sigma\to\infty$ (no-slip wall), with $y$ in units of $\epsilon$. Solid (blue), dashed (orange) and dash-dotted (green) lines: $\dot{x}=0$, $\dot{y}=0$ and $\dot{\theta}=0$; filled circles: equilibria; thick line in (c): a line of equilibria. For $\sigma<\sigma_c$, $(y_y,\pi/2)$ is an attracting node; at $\sigma_c$ the line $y=y_y=y_\theta=\epsilon/\sqrt{2}$ consists of equilibria; for $\sigma>\sigma_c$, orbits above $y_y$ are closed about the centres $(y_\theta,\pm\pi/4)$. Panels (d)--(f) differ in the position of $y_x$ relative to the orbits, which determines whether they loop (Fig.~\ref{fig:regime}). Insets in (d) and (e) magnify $0\le\theta\le\pi/2$, $0.72\le y/\epsilon\le0.95$ and $0.79\le y/\epsilon\le0.935$.}
\label{fig:phase}
\end{figure}

Equilibria of the $(y,\theta)$ subsystem require $\dot {y}=\dot{\theta}=0$. Since $\dot{y}$ vanishes for $\cos{2\theta}=0$ or $y=y_y$, and $\dot{\theta}$ for $\sin{2\theta}=0$ or $y=y_\theta$, the equilibria (modulo $\pi$ in $\theta$) are $(y_y,0)$, $(y_y,\pi/2)$ and $(y_\theta,\pm\pi/4)$. Their linear stability follows from the Jacobian of \eqref{eq:ydot}--\eqref{eq:thdot} (details are given in Appendix~\ref{app:stability}; phase portraits are shown in Fig.~\ref{fig:phase}):
\begin{itemize}
\item $(y_y,\pi/2)$, with the axis normal to the interface, is a stable node for $\sigma<\sigma_c$ and a saddle for $\sigma>\sigma_c$;
\item $(y_y,0)$, with the swimmer axis parallel to the interface, is an unstable node for $\sigma<\sigma_c$ and a saddle for $\sigma>\sigma_c$;
\item $(y_\theta,\pm\pi/4)$ is a saddle for $\sigma<\sigma_c$ and a linear centre for $\sigma>\sigma_c$.
\end{itemize}
At $\sigma=\sigma_c$ the heights coincide, 
\begin{equation}
y_y=y_\theta=y^*=\frac{\epsilon}{\sqrt2},
\end{equation}
and \eqref{eq:ydot}--\eqref{eq:thdot} share the common factor $1-y^{*2}/y^2$. The line $y=y^*$ then consists entirely of equilibria, and away from it the paths in the $(y,\theta)$ plane coincide with those of the pure stresslet discussed at the beginning of this section, $\sin{2\theta}/y = \mathrm{const}$, traversed at a rate rescaled by $1-y^{*2}/y^2$, in the same direction above the line and in the opposite direction below it. The exchange of stability between the equilibria described above takes place through this degenerate configuration (Fig.~\ref{fig:phase}c).

\subsection{Trapping for $\sigma<\sigma_c$}\label{sec:trapping}

\begin{figure}[t]
\centering
\includegraphics[width=\textwidth]{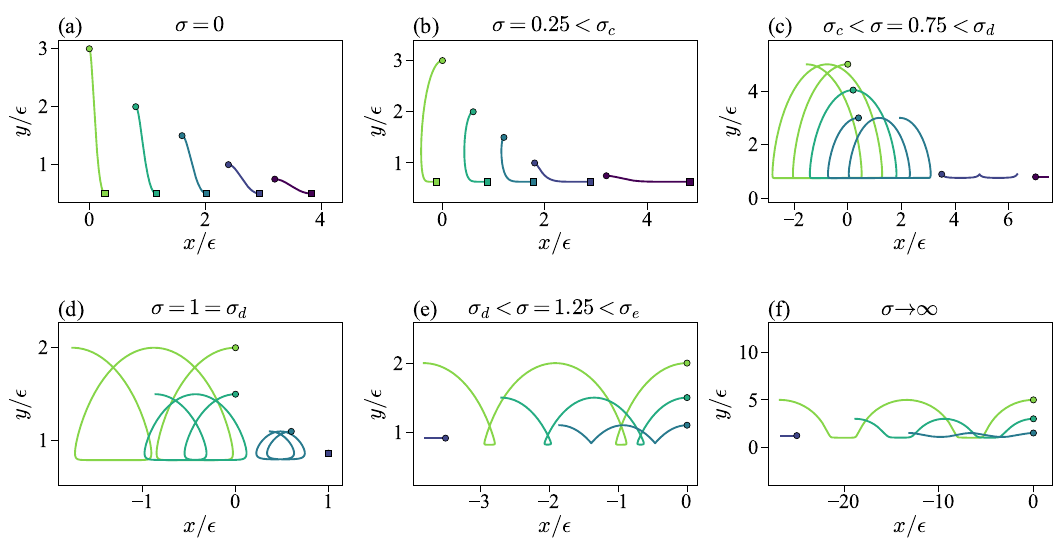}
\caption{Trajectories near a fluid--fluid interface with initial angle $\theta_0=\pi/4$ for (a)~$\sigma=0$, (b)~$0.25$, (c)~$0.75$, (d)~$\sigma_d=1$, (e)~$1.25$ and (f)~$\sigma\to\infty$. Circles: starting points; squares: rest points. Lengths are scaled by $\epsilon$, with equal scales in $x$ and $y$; starting positions $x(0)$ are shifted for clarity. In (a, b) swimmers stop at $y=y_y$. In (c)--(f) periodic orbits are shown over two periods; the swimmer starting at $y_\theta$ glides in a straight line, or is at rest in (d). The orbit from $y_0\approx4.04$ in (c) has zero mean drift (a bound state).}
\label{fig:traj}
\end{figure}

For $\sigma<\sigma_c$, trajectories starting above the interface are attracted to $(y_y,\pi/2)$ (Fig.~\ref{fig:phase}a,b): the swimmer settles at a fixed height with its axis normal to the interface and, since $\dot x\propto\sin{2\theta}\to0$, comes to rest (Fig.~\ref{fig:traj}a,b). This regime includes the free surface ($\sigma=0$), near which the swimmer is trapped at $y_y=\epsilon/2$; there $y_x$ is infinite, so the velocity along the boundary never changes sign. For $0<\sigma<\sigma_c$, whether the swimmer reverses its direction of motion along the boundary before stopping depends on its initial state: it reverses if and only if its trajectory is ever above $y_x$. A no-slip wall does not produce such a state in this model, where the corresponding orbits are periodic. Trapping is therefore induced by the viscosity contrast: it occurs whenever the lower fluid is less than half as viscous as the swimmer's fluid.

Fulford and Blake~\cite{fulford1983motion} suggested, from the viscosity-ratio dependence of the rotation of a slender body near an interface, that a capture region might exist for appropriate viscosity ratios; here such capture appears explicitly. In far-field models the stable orientation depends on the sign of the force dipole: near a no-slip wall pushers align parallel to the wall and pullers perpendicular to it~\cite{berke2008hydrodynamic}, and pushers also align parallel to clean, slipping and surfactant-laden interfaces~\cite{lopez2014dynamics}. The stresslet alone already rotates the swimmer towards the normal orientation; the role of the quadrupole is to turn it into a stable equilibrium at a finite height, which the stresslet alone does not provide. It is reminiscent of the surface-bound state of the bacterium \textit{Thiovulum majus}, which stops swimming and orients normal to the surface while remaining laterally mobile~\cite{petroff2015fast,petroff2018nucleation,das2019transition}, although the mechanism there is different.

Although trapped trajectories end at a node, they still lie on level curves of the first integral. For $\sigma<\sigma_c$ the exponent in \eqref{eq:calH} is positive, so $\mathcal{H}(y_y)=0$ and every curve $\sin2\theta=H_0\mathcal{H}(y)$ passes through the two nodes $(y_y,0)$ and $(y_y,\pi/2)$, where $H(y,\theta)$ is undefined; this is how a conserved quantity is compatible with an attracting and a repelling equilibrium. The first integral therefore also gives the total distance $x_\infty-x(0)$ a trapped swimmer slides before coming to rest, where $x_\infty=\lim_{t\to\infty}x(t)$. Along a trajectory starting from $(y_0,\theta_0)$ with $y_0>y_y$ and $0<\theta_0<\pi/2$, $\sin{2\theta}=H_0\mathcal{H}(y)$ with $H_0=\sin{2\theta_0}/\mathcal{H}(y_0)$, where $\mathcal{H}(y)$ is defined in~\eqref{eq:calH}. By the symmetry $(x,\theta)\mapsto(-x,-\theta)$, starts with $-\pi/2<\theta_0<0$ give the opposite displacement, and on the invariant lines $\sin{2\theta}=0$ there is no motion along the boundary, so it suffices to consider $0<\theta_0<\pi/2$. Eliminating $\theta$, we find
\begin{align}
x_\infty-x(0)&=-\frac{\sigma}{1+\sigma}\left[\int_{y_0}^{y_{\max}}K(y)\,\dd y+\int_{y_y}^{y_{\max}}K(y)\,\dd y\right],
\label{eq:trap_disp}\\
K(y)&=\frac{H_0\mathcal{H}(y)}{\sqrt{1-H_0^2\mathcal{H}^2(y)}}\,\frac{1-y_x^2/y^2}{1-y_y^2/y^2},
\label{eq:Kdef}
\end{align}
for $\theta_0<\pi/4$, where $H_0\mathcal{H}(y_{\max})=1$; for $\theta_0\ge\pi/4$ the swimmer descends directly and the bracket is replaced by $\int_{y_y}^{y_0}K(y)\,\dd y$. Near the node $K(y)\propto(y-y_y)^{\alpha-1}$ with $\alpha=(1-2\sigma)/[2(1+4\sigma)]>0$, so the displacement is finite, but it grows without bound as $\sigma\to\sigma_c^-$ ($\alpha\to0$): just below the transition the swimmer slides ever further before stopping, and at $\sigma_c$ it never stops (Section~\ref{sec:gliding}).

\subsection{Orientation-controlled gliding at $\sigma_c$}\label{sec:gliding}

At $\sigma=\sigma_c=1/2$ the heights $y_y$ and $y_\theta$ coincide at $y^*=\epsilon/\sqrt2$, and both $\dot{y}$ and $\dot{\theta}$ vanish on the line $y=y^*$ for every orientation $\theta$ (Fig.~\ref{fig:phase}c). On this line the swimmer neither approaches nor leaves the interface and does not rotate, but translates parallel to it with speed
\begin{equation}
\dot{x}=\frac{\sqrt2}{3\epsilon}\sin{2\theta}.
\label{eq:glide}
\end{equation}
The speed is set by the orientation alone: it is maximal at $\theta=\pm\pi/4$, vanishes at $\theta=0$ and $\pi/2$, and changes direction under $\theta\mapsto-\theta$. 

\begin{figure}[t]
\centering
\includegraphics[width=0.6\textwidth]{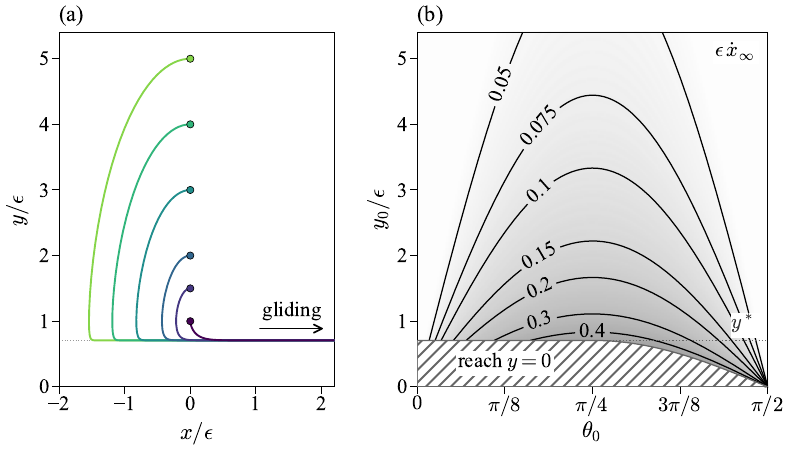}
\caption{Orientation-controlled gliding at $\sigma=\sigma_c=1/2$. (a)~Trajectories in the physical plane, starting from $x(0)=0$ with $\theta_0=\pi/4$ at heights $y_0/\epsilon=1$, $1.5$, $2$, $3$, $4$ and $5$ (dots). All are captured onto the line $y=y^*=\epsilon/\sqrt{2}$ (dotted) and glide in the $+x$ direction at constant speed $\dot{x}_\infty$~\eqref{eq:glide_terminal}. (b)~Terminal speed $\epsilon\,\dot x_\infty=\sin{2\theta_0}/(3y_0/\epsilon)$ as a function of the initial angle and height.}
\label{fig:gliding}
\end{figure}

This gliding state is not a fine-tuned initial condition. Linearising \eqref{eq:ydot}--\eqref{eq:thdot} about a point $(y^*,\theta)$ of the line gives eigenvalues $0$ (along the line) and $2\cos{2\theta}/y^{*2}$, so the part of the line with $\cos{2\theta}<0$, that is $\pi/4<|\theta|<3\pi/4$, is normally attracting. Above the line, every trajectory follows a first-integral curve~\eqref{eq:firstintegral}, $\sin{2\theta}/y=\mathrm{const}$. Taking $0<\theta_0<\pi/2$, the swimmer rotates monotonically towards $\theta=\pi/2$; it first rises while $\theta<\pi/4$, then descends, and reaches the line with an orientation in $\pi/4<\theta<\pi/2$, where the line is attracting. It then glides at constant height and orientation (Fig.~\ref{fig:gliding}a). Since $H(y,\theta)=\sin{2\theta}/y$ is conserved along the approach to the line, a swimmer starting from $(y_0,\theta_0)$ with $y_0>y^*$ glides with terminal orientation $\sin{2\theta_\infty}=y^*\sin{2\theta_0}/y_0$ and terminal speed
\begin{equation}
\dot x_\infty=\frac{\sin{2\theta_0}}{3y_0},
\label{eq:glide_terminal}
\end{equation}
which is independent of $\epsilon$ at fixed stresslet strength. The same expression holds for starts below the line with $\theta_0>\pi/4$ and $y_0>y^*\sin{2\theta_0}$, which rise onto the attracting part of the line; all other starts below $y^*$ reach the boundary. Figure~\ref{fig:gliding}b maps $\dot{x}_\infty$ over initial conditions: its contours are the level curves of $\sin{2\theta}/y$, i.e.\ the trajectories of Fig.~\ref{fig:phase}c.

\subsection{Periodic orbits for $\sigma>\sigma_c$}\label{sec:orbits}

By the symmetry $(x,\theta)\mapsto(-x,-\theta)$ of \eqref{eq:ode}, orbits with $\theta\in(-\pi/2,0)$ are mirror images of those with $\theta\in(0,\pi/2)$, with the same period and opposite drift along the boundary, so it suffices to consider $\theta\in(0,\pi/2)$. For $\sigma>\sigma_c$ and fixed $\theta\in(0,\pi/2)$, the first integral~\eqref{eq:firstintegral} vanishes as $y\to y_y^+$ and as $y\to\infty$ and has a single maximum at $y=y_\theta$, so $0<H(y,\theta)<H_{\max}=H(y_\theta,\pi/4)$, with
\begin{equation}
H_{\max}=\frac{(1+\sigma)^{1/2}}{\epsilon}\left[\frac{2^{1+\sigma}}{3^{3\sigma}}\frac{(2\sigma-1)^{(2\sigma-1)/2}}{\sigma^{3\sigma}}\right]^{1/(1+4\sigma)} .
\label{eq:Hmax}
\end{equation}
Each level set $0<H(y,\theta)<H_{\max}$ is a closed orbit encircling the centre $(y_\theta,\pi/4)$ (Fig.~\ref{fig:phase}d--f): the swimmer oscillates periodically in height and orientation, with larger value of $H(y,\theta)$ corresponding to smaller oscillations. The $(y,\theta)$ subsystem is also reversible under $(t,\theta)\mapsto(-t,\pi/2-\theta)$, a reflection about the line $\theta=\pi/4$ that contains the linear centre, which confirms that $(y_\theta,\pi/4)$ is a nonlinear centre. These periodic orbits form a continuous family; no orbit is selected, and the oscillation amplitude is set by the initial conditions. They generalise the periodic near-wall motion of the treadmiller~\cite{crowdy2010two,crowdy2011treadmilling} to arbitrary viscosity ratio.

In the physical plane, each periodic orbit produces a mean drift along the interface. Its sign structure is controlled by $y_x$, because $\dot x\propto-\sin2\theta\,(1-y_x^2/y^2)$ and $\sin2\theta$ does not change sign along an orbit (the lines $\sin{2\theta}=0$ are invariant). If an orbit crosses $y=y_x$, then $\dot{x}$ changes sign during each period and the swimmer moves backwards for part of the cycle, producing a looping trajectory (Fig.~\ref{fig:traj}c--e). If an orbit stays entirely on one side of $y_x$, $\dot{x}$ keeps one sign and the trajectory is wavy (Fig.~\ref{fig:traj}c,e,f). Orbits fill the region $y>y_y$, with minimum heights arbitrarily close to $y_y$, so looping orbits exist if and only if $y_y<y_x$, that is for $\sigma_c<\sigma<\sigma_e$. For $\sigma>\sigma_e$, which includes the no-slip wall, only wavy trajectories occur (Fig.~\ref{fig:traj}f), consistent with \cite{crowdy2010two,or2009dynamics}. For $\sigma_c<\sigma<\sigma_e$ both looping and wavy types occur, depending on the initial conditions, except at $\sigma=\sigma_d$, where every orbit crosses $y_x=y_\theta$ and loops (Fig.~\ref{fig:traj}d). Looping orbits have been found for a swimmer near a gap in a wall subject to shear~\cite{crowdy2011hydrodynamic}; here they are produced by the viscosity contrast alone, without a background flow.

\subsection{Mean drift, drift reversal and bound states}\label{sec:bound}

The mean drift velocity of an orbit with $H(y,\theta)=H_0\in(0,H_{\max})$ is $\langle\dot x\rangle=\Delta x/T$, where $T$ is the period and $\Delta x$ the displacement per period. Using the first integral to eliminate $\theta$, both reduce to quadratures over the height,
\begin{align}
T&=2\int_{y_{\min}}^{y_{\max}}\frac{y}{\sqrt{1-H_0^2\mathcal{H}^2(y)}}\left(1-\frac{y_y^2}{y^2}\right)^{-1}\dd y,\label{eq:T}\\
\Delta x&=-\frac{2\sigma}{1+\sigma}\int_{y_{\min}}^{y_{\max}}K(y)\,\dd y,\label{eq:Dx}
\end{align}
where $\mathcal{H}(y)$ and $K(y)$ are defined in \eqref{eq:calH} and \eqref{eq:Kdef}, respectively, and $y_{\min}<y_{\max}$ are the roots of $H_0\mathcal{H}(y)=1$. Thus $\Delta x$ has the same integrand as the slide distance~\eqref{eq:trap_disp} of a trapped swimmer, taken around a closed orbit instead of along a trajectory ending at the node. Because the first integral eliminates $\theta$, $T$ and $\Delta x$ are one-dimensional integrals over $y$ that can be evaluated accurately for every orbit. Computing them instead by integrating trajectories in time is costly for large orbits, whose periods grow as $T\sim2/H_0^2$ for small $H_0$ (see below), and over such long times numerical errors in $\Delta x$ accumulate.

The two ends of the family of orbits can be treated asymptotically. For large orbits, $H_0\to0^+$, we have $y_{\min}\to y_y$, $y_{\max}\sim1/H_0$, $T\sim2/H_0^2$ and $\Delta x\sim-2\sigma/[(1+\sigma)H_0]$, so that
\begin{equation}
\langle\dot{x}\rangle\sim-\frac{\sigma}{1+\sigma}\,H_0\to0^- \qquad (H_0\to0^+),
\label{eq:drift_small}
\end{equation}
for $\theta\in(0,\pi/2)$. For small orbits, $H_0\to H_{\max}$, the drift tends to the velocity parallel to the boundary at the centre,
\begin{equation}
\langle\dot{x}\rangle\to\dot x(y_\theta,\pi/4)=\frac{1}{\epsilon}\sqrt{\frac{2\sigma}{3(1+\sigma)}}\,\frac{4\sigma+1}{6\sigma^2}(1-\sigma)\qquad (H_0\to H_{\max}).
\label{eq:drift_centre}
\end{equation}
The drift at the centre therefore reverses direction at $\sigma=\sigma_d=1$: for $\sigma<1$ small orbits drift in the positive direction, opposite to large orbits, while for $\sigma>1$ all orbits drift the same way. A drift reversal driven by a higher-order singularity echoes the hydrodynamic mobility reversal found for squirmers near surfaces~\cite{kuron2019hydrodynamic}, and the dependence on viscosity ratio has passive analogues: a sphere translating parallel to a fluid interface can rotate either with or against the sense of rolling, depending on its distance from the interface and the viscosity ratio~\cite{lee1980motion}, and a slender body oriented normal to the interface changes its sense of rotation with distance for viscosity ratios between $2/3$ and about $0.77$~\cite{fulford1983motion}.

\begin{figure}[t]
\centering
\includegraphics[width=0.75\textwidth]{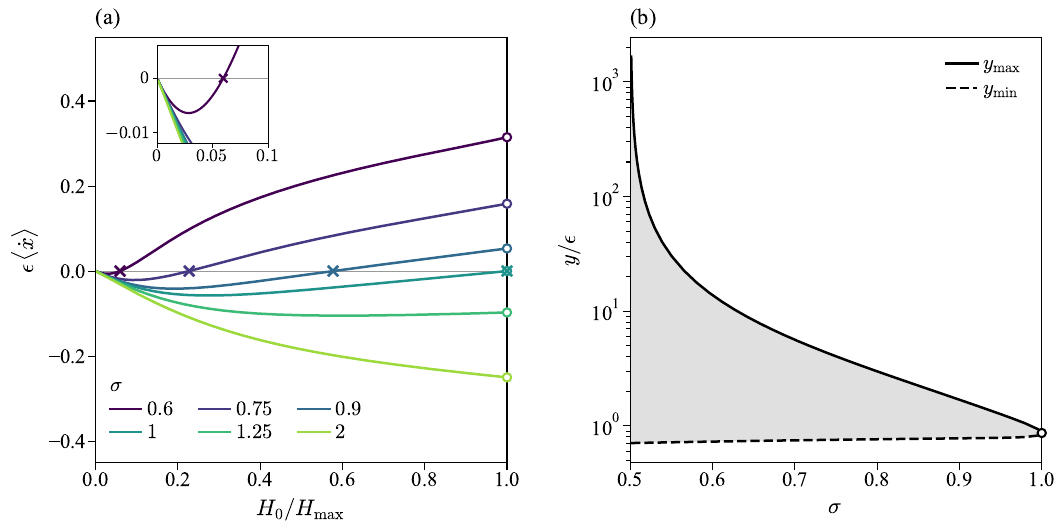}
\caption{Mean drift and bound states near a fluid--fluid interface, from \eqref{eq:T}--\eqref{eq:Dx} with $\theta\in(0,\pi/2)$. (a)~$\epsilon\langle\dot{x}\rangle$ against $H_0/H_{\max}$; open circles, centre value \eqref{eq:drift_centre}; crosses, bound states. Inset: magnification near $H_0/H_{\max}=0$, where the drift is negative for all $\sigma$. (b)~Height range of the bound orbit for $\sigma_c<\sigma<\sigma_d$; it diverges as $\sigma\to\sigma_c^+$ and shrinks onto the centre at $\sigma=\sigma_d$ (open circle).}
\label{fig:bound}
\end{figure}

For $\sigma_c<\sigma<\sigma_d$, \eqref{eq:drift_small} and \eqref{eq:drift_centre} have opposite signs, so by continuity at least one orbit has $\langle\dot{x}\rangle=0$. This is a hydrodynamic bound state: the swimmer oscillates indefinitely with zero net displacement. Numerical evaluation of \eqref{eq:T}--\eqref{eq:Dx} shows that this orbit is unique (Fig.~\ref{fig:bound}a). Its height range is shown in Fig.~\ref{fig:bound}b: the bound orbit grows without bound as $\sigma\to\sigma_c^+$, mirroring the divergence of the slide distance as $\sigma\to\sigma_c^-$ (Section~\ref{sec:trapping}), and shrinks onto the centre as $\sigma\to\sigma_d^-$. At $\sigma=\sigma_d$ the centre becomes a true equilibrium of the full system~\eqref{eq:ode}, with $\dot{x}=\dot{y}=\dot{\theta}=0$, while every other orbit drifts. For $\sigma_d<\sigma<\sigma_e$ we find $\Delta x<0$ for all orbits, and for $\sigma\ge\sigma_e$ no orbit loops, so no bound state exists. Non-trivial bound states therefore occur only in the window
\begin{equation}
\frac{1}{2}<\sigma<1,
\label{eq:window}
\end{equation}
that is, when the lower fluid is less viscous than the swimmer's fluid but by no more than a factor of two. Hydrodynamic bound states have been found for a swimmer near a gap in a wall~\cite{crowdy2011hydrodynamic}, for a treadmiller in a corner~\cite{davis2012stresslet} and between pairs of swimming algae~\cite{drescher2009dancing}; here they arise from a uniform flat interface.

\section{Swimmer near a partial-slip wall}\label{sec:slip}

Following Lopez and Lauga~\cite{lopez2014dynamics} who investigated a swimmer's dynamics near a two-fluid interface and a partial-slip wall in the three-dimensional case, it is of interest to compare the results just obtained with the dynamics of the same swimmer in a single fluid with a boundary at $y=0$ where a Navier-slip condition holds. It is found below that the slip wall shares the three-regime structure of the interface, with a transition at a critical slip length, but the secondary crossings of the characteristic heights occur in the reverse order, on the trapping side of the transition: the slip wall has no looping trajectories or bound states, but has a range of slip lengths in which no trapped swimmer reverses its direction of motion.

As in Section~\ref{sec:interface}, we first remove the source quadrupole. Setting $\tilde\epsilon=0$ in~\eqref{eq:ode_slip}, the stresslet obeys
\begin{equation}
\dot{x}=-s_x(\tilde{y})\,\frac{\sin{2\theta}}{y},\qquad \dot{y}=s_y(\tilde{y})\,\frac{\cos{2\theta}}{y},\qquad \dot{\theta}=s_\theta(\tilde{y})\,\frac{\sin{2\theta}}{2y^2},
\label{eq:stresslet_slip}
\end{equation}
which has the form of~\eqref{eq:ode_stresslet} with the constants $s_x=\sigma/(1+\sigma)$ and $s_y=s_\theta=1$ of the interface replaced by functions of $\tilde y=y/\lambda$ alone (Appendix~\ref{app:stresslet_slip}). These satisfy $0<s_x(\tilde{y})<1$ and $1/2<s_y(\tilde{y})\le1$ at all heights, with $s_\theta(\tilde{y})=2s_y(\tilde{y})-1$, so the signs of $\dot{x}$, $\dot{y}$ and $\dot{\theta}$ are those found for the interface. The stresslet therefore again rotates towards the normal orientation and reaches the wall in finite time, without equilibria or reversal of the drift: as for the interface, trapping, gliding and periodic orbits require the source quadrupole. The difference is that the slip length introduces a length scale. The drift factor $s_x(\tilde{y})$ rises monotonically from $0$ for $\tilde{y}\ll1$ to $1$ for $\tilde{y}\gg1$, so the wall acts on the drift like an interface with an effective viscosity ratio
\begin{equation}
\sigma_{\mathrm{eff}}(\tilde{y})=\frac{s_x(\tilde{y})}{1-s_x(\tilde{y})},
\end{equation}
which decreases from $\infty$ (no-slip) far from the wall to $0$ (free surface) close to it. The approach and rotation rates are also reduced at heights comparable to $\lambda$ (Appendix~\ref{app:stresslet_slip}), so that, unlike for the interface, the paths of the stresslet in the $(y,\theta)$ plane depend on the boundary.

\subsection{Equilibria and the critical slip length}\label{sec:slip_equilibria}

\begin{figure}[t]
\centering
\includegraphics[width=0.6\textwidth]{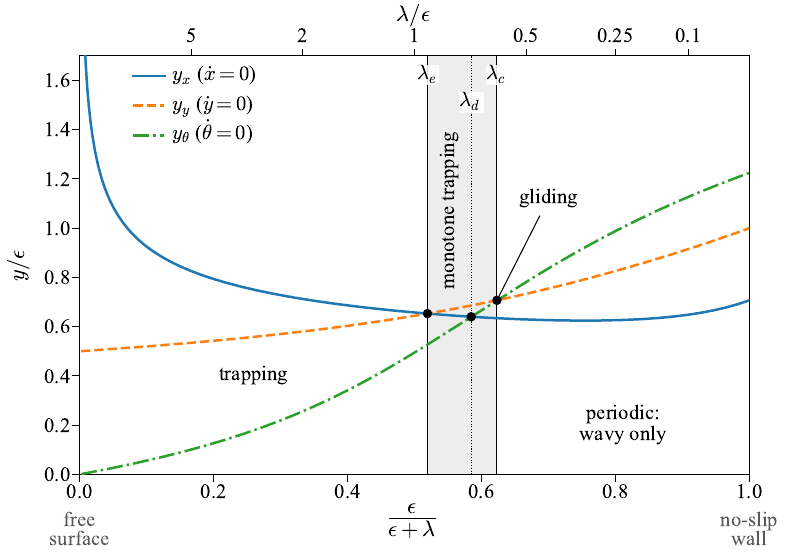}
\caption{Regime diagram for the swimmer near a partial-slip wall. The heights $y_x$, $y_y$ and $y_\theta$ at which $\mathcal{F}_x(\tilde{y};\tilde{\epsilon})$, $\mathcal{F}_y(\tilde{y};\tilde{\epsilon})$ and $\mathcal{F}_\theta(\tilde{y};\tilde{\epsilon})$ of \eqref{eq:Fxslip}--\eqref{eq:Fthetaslip} vanish, in units of $\epsilon$, are plotted against $\epsilon/(\epsilon+\lambda)$, which runs from the free surface ($0$) to the no-slip wall ($1$); values of $\lambda/\epsilon$ are shown on the upper axis. The crossings $y_y=y_\theta$, $y_x=y_\theta$ and $y_x=y_y$ (dots) define $\lambda_c\approx0.605\,\epsilon$, $\lambda_d\approx0.711\,\epsilon$ and $\lambda_e\approx0.926\,\epsilon$; compared with Fig.~\ref{fig:regime}, they occur in the reverse order, all on the trapping side of the transition. Labels indicate the long-time behaviour in each regime.}
\label{fig:regime_slip}
\end{figure}

\begin{figure}[t]
\centering
\includegraphics[width=\textwidth]{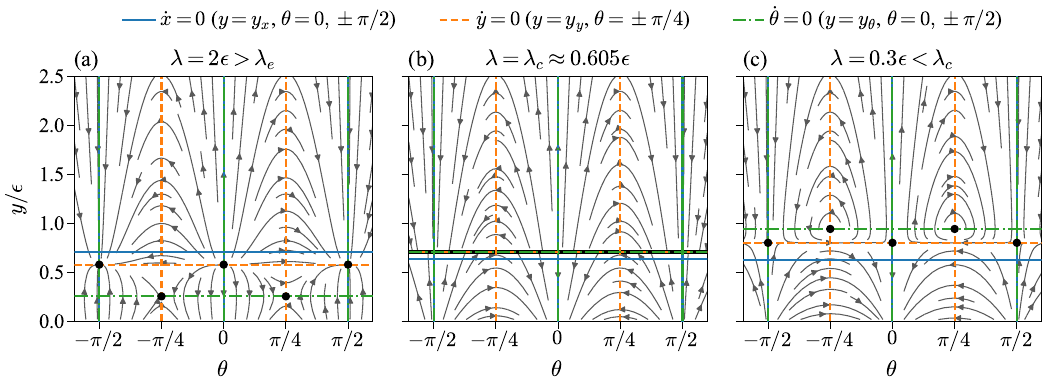}
\caption{Phase portraits of the $(y,\theta)$ subsystem near a partial-slip wall for (a)~$\lambda=2\epsilon$, (b)~$\lambda=\lambda_c\approx0.605\epsilon$ and (c)~$\lambda=0.3\epsilon$. Solid (blue), dashed (orange) and dash-dotted (green) lines: $\dot{x}=0$, $\dot{y}=0$ and $\dot{\theta}=0$; circles: equilibria. In (b) the line $y=y_y=y_\theta=\epsilon/\sqrt{2}$ (thick) consists of equilibria.}
\label{fig:slip}
\end{figure}

\begin{figure}[t]
\centering
\centering
\includegraphics[width=\textwidth]{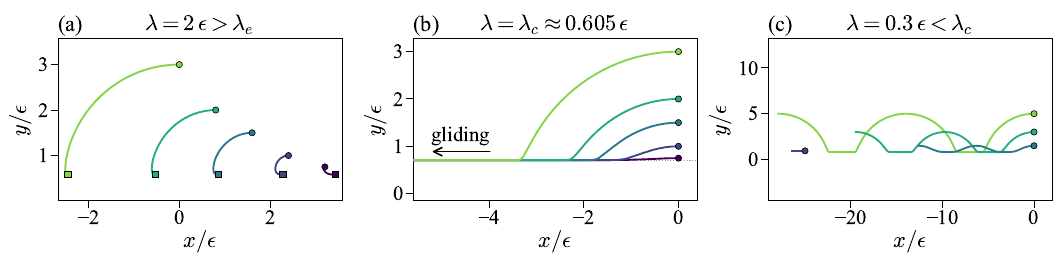}
\caption{Trajectories near a partial-slip wall for (a)~$\lambda=2\epsilon$, (b)~$\lambda=\lambda_c$ and (c)~$\lambda=0.3\epsilon$, all with initial angle $\theta_0=\pi/4$. Circles: starting points; squares: rest points. Lengths are scaled by $\epsilon$, with equal scales in $x$ and $y$; starting positions $x(0)$ are shifted for clarity. In (a) swimmers stop at $y=y_y$; in (b) they are captured onto $y=y^*=\epsilon/\sqrt{2}$ and glide in the $-x$ direction; in (c) periodic orbits are shown over two periods, and the swimmer starting at $y_\theta$ glides in a straight line.}
\label{fig:traj_slip}
\end{figure}

The equations~\eqref{eq:ode_slip} share the angular structure of \eqref{eq:ode}. Equilibria with $\cos2\theta\neq0$ lie at a positive root $\tilde{y}_y$ of $\mathcal{F}_y(\tilde{y};\tilde{\epsilon})=0$, and those with $\sin{2\theta}\neq0$ at a root $\tilde{y}_\theta$ of $\mathcal F_\theta(\tilde y;\tilde\epsilon)=0$; we write $y_y=\lambda\tilde{y}_y$ and $y_\theta=\lambda\tilde{y}_\theta$ for the corresponding heights. From \eqref{eq:Fyslip}--\eqref{eq:Fthetaslip}, $\mathcal{F}_y(\tilde{y};\tilde{\epsilon})$ and $\mathcal{F}_\theta(\tilde{y};\tilde{\epsilon})$ both diverge to $-\infty$ as $\tilde{y}\to0^+$ when $\tilde{\epsilon}\neq0$, while $\mathcal{F}_y(\tilde{y};\tilde{\epsilon})\sim2/\tilde{y}$ and $\mathcal{F}_\theta(\tilde{y};\tilde{\epsilon})\sim1/\tilde y^2$ as $\tilde{y}\to\infty$, so each has at least one positive root; numerically we find that each root is unique for $0.002\le\lambda/\epsilon\le100$ (Fig.~\ref{fig:regime_slip}). We find that the two roots can coincide only at $y=\epsilon/\sqrt{2}$, so the critical value is the root of $\mathcal F_\theta(\tilde\epsilon/\sqrt2;\tilde\epsilon)=0$ (Appendix~\ref{app:nullcline}):
\begin{equation}
\tilde\epsilon_c=\epsilon/\lambda_c\approx1.653,\qquad\text{i.e.}\qquad \lambda_c\approx0.605\,\epsilon .
\label{eq:lambdac}
\end{equation}
This crossing plays exactly the role of $\sigma_c$. The equilibria are classified as for the interface (Appendix~\ref{app:stability_eq}):
\begin{itemize}
\item $(y_y,\pi/2)$, with the axis normal to the wall, is a stable node for $\lambda>\lambda_c$ and a saddle for $\lambda<\lambda_c$;
\item $(y_y,0)$ is an unstable node for $\lambda>\lambda_c$ and a saddle for $\lambda<\lambda_c$;
\item $(y_\theta,\pm\pi/4)$ is a saddle for $\lambda>\lambda_c$ and a linear centre for $\lambda<\lambda_c$.
\end{itemize}
For $\lambda<\lambda_c$ the phase portrait therefore resembles that of the interface with $\sigma>\sigma_c$, with periodic orbits about the centre, and for $\lambda>\lambda_c$ the swimmer is trapped with its axis normal to the wall (Figs.~\ref{fig:slip} and~\ref{fig:traj_slip}). At $\lambda=\lambda_c$ the line $y=\epsilon/\sqrt{2}$ consists of equilibria and, as at $\sigma_c$, the swimmer glides along the wall at a speed set by its orientation (Fig.~\ref{fig:traj_slip}b), but in the opposite direction (cf.~Fig.~\ref{fig:gliding}a): here the line lies above $y_x$, whereas at the interface it lies below $y_x=\epsilon$. 

\subsection{Trapping for $\lambda>\lambda_c$}\label{sec:slip_trapping}

For $\lambda>\lambda_c$, as at the interface for $\sigma<\sigma_c$, trapping is guaranteed for swimmers starting above $y_y$: the line $y=y_y$ is invariant, since $\dot{y}=0$ on it for every $\theta$, so such a swimmer remains above it and, by the classification of Section~\ref{sec:slip_equilibria}, comes to rest at $(y_y,\pi/2)$ with its axis normal to the wall (Figs.~\ref{fig:slip}a and~\ref{fig:traj_slip}a).

How it comes to rest depends on the height $y_x=\lambda\tilde{y}_x$ at which the velocity parallel to the wall vanishes, $\tilde{y}_x$ being the positive root of $\mathcal{F}_x(\tilde{y};\tilde{\epsilon})=0$. Numerically, $y_x$ crosses $y_y$ at $\lambda_e\approx0.926\,\epsilon$ (Fig.~\ref{fig:regime_slip}). For $\lambda>\lambda_e$, $y_x>y_y$, so the final approach to the node takes place below $y_x$, and a swimmer that has been above $y_x$ reverses its direction of motion along the wall before stopping. This is also the behaviour at the interface, where $y_x>y_y$ for all $\sigma<\sigma_c$. For $\lambda_c<\lambda<\lambda_e$, by contrast, $y_x<y_y$, so the velocity parallel to the wall has a single sign throughout $y>y_y$: the swimmer slides in one direction and stops without reversing. This monotone trapping has no counterpart at the interface. The crossing $y_x=y_\theta$ at $\lambda_d\approx0.711\,\epsilon$ also lies in this regime, but since $y_\theta<y_y$ there it affects only swimmers starting below $y_y$.

Swimmers starting below $y_y$ either rise to the node or reach the wall, the two outcomes being separated by the stable manifold of the saddle $(y_\theta,\pi/4)$. The trapping height satisfies $y_y<\epsilon/\sqrt2$ for all $\lambda>\lambda_c$.

\subsection{Periodic orbits for $\lambda<\lambda_c$}\label{sec:slip_orbits}

Because \eqref{eq:ode_slip} has the same angular structure as \eqref{eq:ode}, it also possesses a first integral (Appendix~\ref{app:firstintegral}). For $\lambda<\lambda_c$ its level sets above $y_y$ are closed orbits about the centre $(y_\theta,\pi/4)$ (Figs.~\ref{fig:slip}c and~\ref{fig:traj_slip}c). As at the interface, the orbits form a continuous family, and the amplitude of the oscillation is set by the initial conditions. 

Whether these orbits loop is again decided by the position of $y_x$ relative to the orbits. For every $\lambda<\lambda_c$ the three heights are ordered as
\begin{equation}
y_x<y_y<y_\theta ,
\label{eq:ordering_slip}
\end{equation}
as shown by the regime diagram (Fig.~\ref{fig:regime_slip}); this is the ordering~\eqref{eq:ordering} of the interface for $\sigma>\sigma_e$. The orbits therefore lie entirely above $y_x$, where $\mathcal{F}_x(\tilde{y};\tilde{\epsilon})>0$, so the velocity along the wall keeps one sign throughout each orbit: every trajectory is wavy and drifts in the $-x$ direction for $\theta\in(0,\pi/2)$, and there are neither looping trajectories nor hydrodynamic bound states. The crossings $y_x=y_\theta$ and $y_x=y_y$ that open these windows at the interface, at $\sigma_d$ and $\sigma_e$, occur near the slip wall at $\lambda_d$ and $\lambda_e$, both of which exceed $\lambda_c$; they lie on the trapping side of the transition, where they instead produce the two kinds of trapping of Section~\ref{sec:slip_trapping}. For every $\lambda<\lambda_c$ the slip wall thus behaves like an interface with $\sigma\ge\sigma_e$.

In other words, wall slip converts the oscillatory near-wall motion of the no-slip problem into trapping once the slip length exceeds about $0.6\epsilon$. This is a two-dimensional, closed-form counterpart of the critical slip lengths reported for squirmers and micromotors, above which swimmers that escape a no-slip wall become trapped or slide stably along it~\cite{poddar2020near,ghosh2026slippery}, or wall-bound sliding and oscillating states appear~\cite{guan2024effect}, and above which oscillatory states in channel flow are destabilised~\cite{ghosh2024harnessing}. In \cite{poddar2020near} a short-range repulsion is included to prevent contact with the wall, although the transitions are shown to be caused by slip; here, trapping is purely hydrodynamic and requires no repulsion. In the present model the critical slip length is set by $\epsilon$, which is consistent with the physical expectation that slip matters once it is comparable to the scale of the near-field flow.

\subsection{Comparison with the fluid--fluid interface}\label{sec:slip_compare}

Table~\ref{tab:regimes} compares the long-time behaviour near the two boundaries. They share the same three-regime structure, with $\lambda<\lambda_c$, $\lambda=\lambda_c$ and $\lambda>\lambda_c$ corresponding to $\sigma>\sigma_c$, $\sigma=\sigma_c$ and $\sigma<\sigma_c$, and their extreme limits coincide exactly. The transition occurs at the same height, $y^*=\epsilon/\sqrt{2}$, for both (Appendix~\ref{app:nullcline}), and in both cases the stresslet alone produces none of these regimes (Sections~\ref{sec:interface} and~\ref{sec:slip}).

\begin{table}[t]
\caption{Long-time behaviour near a fluid--fluid interface and a partial-slip wall, ordered from the free-surface limit (top) to the no-slip limit (bottom); $\sigma_c=1/2$, $\sigma_d=1$, $\sigma_e=(1+\sqrt{3})/2$, $\lambda_c\approx0.605\,\epsilon$, $\lambda_e\approx0.926\,\epsilon$. The slip-wall crossing $y_x=y_\theta$ at $\lambda_d\approx0.711\,\epsilon$ affects only swimmers starting below $y_y$ and is not listed.}
\label{tab:regimes}
\small
\begin{tabular}{@{}p{0.21\textwidth}p{0.36\textwidth}p{0.34\textwidth}@{}}
\toprule
$(y,\theta)$ dynamics & Fluid--fluid interface & Partial-slip wall\\
\midrule
attracting node at $(y_y,\pi/2)$
 & $\sigma<\sigma_c$: trapped; reverses if it has been above $y_x$
 & $\lambda>\lambda_e$: trapped; reverses if it has been above $y_x$\\ \addlinespace[2pt]
 & & $\lambda_c<\lambda\le\lambda_e$: trapped without reversal\\
\midrule
line of equilibria $y=\epsilon/\sqrt2$
 & $\sigma=\sigma_c$: glides, speed $\propto\sin2\theta$
 & $\lambda=\lambda_c$: glides in the opposite direction\\
\midrule
periodic orbits about $(y_\theta,\pm\pi/4)$
 & $\sigma_c<\sigma<\sigma_d$: wavy or looping; unique bound state
 & $\lambda<\lambda_c$: wavy only; no bound state\\ \addlinespace[2pt]
 & $\sigma=\sigma_d$: looping only; centre stationary & \\ \addlinespace[2pt]
 & $\sigma_d<\sigma<\sigma_e$: wavy or looping; no bound state & \\ \addlinespace[2pt]
 & $\sigma\ge\sigma_e$: wavy only; no bound state & \\
\bottomrule
\end{tabular}
\end{table}

The two boundaries differ in how the remaining crossings are arranged. At the interface, the crossings $y_x=y_\theta$ and $y_x=y_y$ occur at $\sigma_d$ and $\sigma_e$, on the oscillatory side of the transition, where they open the windows of looping trajectories and bound states. Near the slip wall the same crossings occur at $\lambda_d$ and $\lambda_e$, on the trapping side, where they instead separate two kinds of trapping (Section~\ref{sec:slip_trapping}). As a result, the slip wall has neither looping trajectories nor bound states, but has a range of slip lengths in which no trapped swimmer reverses, and at the transition the swimmer glides in the opposite direction.

The difference originates in the boundary conditions. At the interface, the conditions relate $h_+(z)$ and $h_-(z)$ algebraically on $y=0$, so the image system is a finite set of point singularities at the mirror point $\zob$, with strengths that depend on $\sigma$ through constant coefficients such as $(1-\sigma)/(1+\sigma)$; this is the two-dimensional analogue of the image systems of \cite{aderogba1978action}. The Navier condition instead relates $h(z)$ to its derivative. Solving the resulting first-order equation for $h(z)$ shows that the images of the force singularities, carried by $f(z)=[-h(z)+\hat{g}'(z)]/2$, consist of the point image for a perfectly slipping wall at $\zob$ together with a line distribution of the reflected singularities of $h(z)$ along the ray extending downwards from $\zob$, with weight $\ee^{-\ell/(2\lambda)}$ at depth $\ell$ below it; the images of the source singularities in $\hat{g}'(z)$ remain at $\zob$ (Appendix~\ref{app:slip_image}). A similar structure was found in three dimensions~\cite{lauga2005brownian} and underlies the far-field analysis of swimmers near slipping walls in \cite{lopez2014dynamics}. The slip length thus sets the depth of a continuous image distribution, which makes the dynamics depend on $y/\lambda$.

Near the no-slip limit this acts, to first order, like a no-slip wall displaced to $y=-\lambda$, so all three heights decrease by the same amount~\eqref{eq:heights_small_slip} and their separations are unchanged; a large but finite viscosity ratio instead raises $y_x$ and lowers $y_y$ and $y_\theta$~\eqref{eq:heights_large_sigma}. Moving away from the no-slip limit, $y_x$ at the interface therefore immediately approaches the other two heights and crosses them before $y_y$ and $y_\theta$ meet at $\sigma_c$, whereas near the slip wall it initially falls with them, reaches a minimum and crosses them only after they have met at $\lambda_c$ (Fig.~\ref{fig:regime_slip}). Near the free-surface limit, the drift along the slip wall contains a term proportional to $(y/\lambda)\log(\lambda/y)$ already at first order, so the approach to that limit is logarithmically slow, whereas the approach as $\sigma\to0$ is algebraic (Appendix~\ref{app:asym}).

Finally, the slip length introduces a length scale, so the critical slip lengths scale with $\epsilon$, whereas the critical viscosity ratios are universal. A partial-slip wall and a viscous interface therefore produce similar phase portraits and the same primary transition, but they are not equivalent boundaries.

\section{Discussion}\label{sec:discussion}

We have derived closed-form equations of motion for a two-dimensional treadmilling swimmer model near a flat interface between fluids of viscosity ratio $\sigma$ and near a partial-slip wall with slip length $\lambda$. For the interface, the dynamics are organised by three characteristic heights whose crossings define universal critical viscosity ratios $\sigma_c=1/2$, $\sigma_d=1$ and $\sigma_e=(1+\sqrt{3})/2$. Below $\sigma_c$ the swimmer is trapped with its axis normal to the interface. At $\sigma_c$ it glides at a speed selected by its orientation, and its terminal speed is predicted exactly by a first integral. Above $\sigma_c$ it follows periodic orbits, and the mean drift of small orbits reverses at $\sigma_d$. Hydrodynamic bound states occur only for $\sigma_c<\sigma<\sigma_d$, and looping trajectories only for $\sigma_c<\sigma<\sigma_e$. Near the partial-slip wall the same primary transition occurs at a critical slip length $\lambda_c\approx0.605\,\epsilon$, beyond which the swimmer is trapped. The secondary crossings, however, occur in the reverse order, on the trapping side of the transition. As a result, the slip wall has no looping trajectories or bound states, but it has a range of slip lengths in which no trapped swimmer reverses. At the transition, the swimmer also glides in the opposite direction. These results provide an explicit dynamical-systems picture of how a single boundary parameter can switch a microswimmer between qualitatively different modes of locomotion.

The swimmer in this study does not self-propel, so its entire motion is the response to its own image flow. At the level of the leading singularity, the stresslet, both boundaries act alike: the stresslet rotates towards the boundary normal and approaches the boundary, and the boundary only modifies the rates, through the constant $\sigma/(1+\sigma)$ for the interface and through functions of $y/\lambda$ for the slip wall (Sections~\ref{sec:interface} and~\ref{sec:slip}). Every regime found here is produced by the next singularity, the source quadrupole. Its contributions to the velocity normal to the boundary and to the rotation rate oppose those of the stresslet close to the boundary, and the heights $y_y$ and $y_\theta$ are where the two balance for translation and rotation, respectively. When translational balance is reached further from the boundary than rotational balance ($\sigma<\sigma_c$, $\lambda>\lambda_c$), the swimmer is held at $y_y$ with its axis normal to the boundary; when the order is reversed, the balance becomes oscillatory; and the gliding state is the borderline at which both balances occur at the same height, $\epsilon/\sqrt{2}$ for both boundaries, for every orientation. The leading singularity thus fixes the generic tendency, while the higher-order one decides the qualitative dynamics and distinguishes the boundaries. A similar division appears in the three-dimensional far-field analysis of \cite{lopez2014dynamics}, where the force dipole produces the same attraction and parallel alignment near all the interfaces considered, while the sense of circular swimming is set by the rotlet dipole and depends on the boundary condition.

Several of the states found here have counterparts in other models, simulations and experiments, although the mechanisms are not always the same. States at a finite height above a boundary have been reported for catalytic particles near a wall~\cite{uspal2015self}, for bacteria above surfaces~\cite{htet2024hydrodynamic} and for squirmers near a liquid--liquid interface~\cite{feng2023dynamics}, where strong pullers swim parallel to the interface at a fixed distance while tilted towards it. In these studies the swimmers are self-propelled, and the hovering height results from a balance between self-propulsion and the boundary-induced flow, supplemented in \cite{uspal2015self} by the phoretic interaction with the wall. Here the swimmer does not self-propel, and the trapped state for $\sigma<\sigma_c$ arises purely from the competition between the images of the stresslet and of the source quadrupole, which balance at $y_y$ for translation and at $y_\theta$ for rotation. The parameter regimes also differ: the hovering state of \cite{feng2023dynamics} occurs for strong pullers in the less viscous fluid, corresponding to $\sigma>1$ in our notation, where our swimmer instead oscillates, whereas our trapped states occur for $\sigma<1/2$, a regime in which swimmers in their simulations mostly cross the penetrable interface. By contrast, experiments with \textit{E.~coli} show stronger accumulation near a gas interface than near liquid and solid boundaries~\cite{ahmadzadegan2019hydrodynamic}, consistent in trend with the trapping found here for $\sigma<\sigma_c$, although the mechanism there also involves self-propulsion, interfacial viscosity and short-range physicochemical interactions. The normal orientation of the trapped state recalls the surface-bound state of \textit{Thiovulum majus}~\cite{petroff2015fast,petroff2018nucleation,das2019transition}, although, as noted in Section~\ref{sec:trapping}, the mechanism there is different.

The gliding state at $\sigma=\sigma_c$ resembles the sliding states found for active particles near walls~\cite{uspal2015self} and near slipping walls~\cite{poddar2020near,ghosh2026slippery}, but differs from them in structure. Sliding states are typically isolated attractors that persist over a range of parameters, whereas the gliding state here is a continuous family, one for each orientation, that exists only at $\sigma=\sigma_c$ and is set by the initial condition through the first integral. The reversal of the mean drift at $\sigma_d$ is driven by the source quadrupole, in line with the mobility reversal found for squirmers, where higher-order flow moments set the direction of motion along a surface~\cite{kuron2019hydrodynamic} (Section~\ref{sec:bound}); here the reversal is controlled by the viscosity ratio alone.

For the partial-slip wall, the change from oscillation to trapping beyond a threshold slip length parallels the slip-induced transitions reported for squirmers, micromotors and elongated swimmers~\cite{poddar2020near,ghosh2026slippery,ghosh2024harnessing,guan2024effect} (Section~\ref{sec:slip_orbits}). Unlike in \cite{poddar2020near}, where a short-range repulsion is included to prevent contact with the wall, trapping here is purely hydrodynamic, and the threshold slip length is set by $\epsilon$, $\lambda_c\approx0.605\,\epsilon$, consistent with the expectation that slip matters once it is comparable to the scale of the near-field flow.

These correspondences are qualitative. The present model shows that such transitions can arise as explicit bifurcations of a low-dimensional dynamical system controlled by a single boundary parameter, and it identifies the mechanism through the ordering of three characteristic heights. More broadly, because a single boundary parameter, the viscosity ratio or the slip length, switches the swimmer between qualitatively different modes of motion, and near a slip wall the threshold also depends on the swimmer size, fluid interfaces and surfaces of controlled or patterned slip might be used to guide or sort microswimmers. Hydrodynamic interactions with flows and boundaries have been proposed and used for sorting swimmers by motility~\cite{mathijssen2016upstream,zaferani2018rheotaxis}, and wall slip has been proposed to direct or probe swimmer motion~\cite{lemelle2013curvature,hu2015physical,ketzetzi2020slip}. We note that a two-dimensional model cannot capture phenomena that rely on the third dimension, such as circular trajectories near walls and free surfaces and their reversal near slipping surfaces~\cite{lauga2006swimming,di2011swimming,lemelle2013curvature,hu2015physical,lopez2014dynamics,lemelle2010counterclockwise}, or the dynamics of chiral swimmers near slip walls~\cite{guan2025study}.

The present model also points to several natural extensions. First, all results here are exact for the point singularity, a stresslet combined with a source quadrupole, but their relevance to a finite swimmer depends on how $\epsilon$ is interpreted. If $\epsilon$ is identified with the radius of the circular treadmiller of Appendix~\ref{app:point_swimmer}, the point-singularity description of that swimmer is expected to be quantitatively accurate only when it is several radii from the boundary, whereas several of the states found here lie at heights below $\epsilon$, where such a swimmer would overlap the boundary. An important next step is therefore to test the transitions against an exact solution for a finite treadmiller near an interface or a slip wall, in the spirit of~\cite{crowdy2011treadmilling}, or to extend the matched-asymptotic corrections of~\cite{davis2013matched} to these boundaries. Such corrections are known to change the predicted long-time behaviour close to a no-slip wall~\cite{davis2013matched,davis2012stresslet}, and it would be valuable to determine how they shift the thresholds $\sigma_c$ and $\lambda_c$. Second, allowing the interface to deform, or the swimmer to cross it, would connect the present results with the additional modes found in simulations~\cite{feng2022dynamics,feng2023dynamics,shaik2017motion}.

The framework itself also extends naturally: the construction of Appendix~\ref{app:interface} applies to any point singularity near a viscosity discontinuity, which would allow image series to be built for swimmers in thin films bounded by two interfaces~\cite{mathijssen2016hydrodynamics}. Squirmer-type swimmers with both self-propulsion and a stresslet, that is pushers and pullers, can be treated within the same framework by adding a source dipole, and curved interfaces, near which trapping depends on the drop size~\cite{spagnolie2015geometric,desai2018hydrodynamics}, may be accessible by similar techniques.

\begin{acknowledgements}
T.K. acknowledges the support of the Toyota Physical and Chemical Research Institute. T.K. also acknowledges support by grant \#63132 from the John Templeton Foundation, as recipient of an Enrico Fermi Fellowship awarded through the Center for SpaceTime and the Quantum. The opinions expressed in this publication are those of the authors and do not necessarily reflect the views of the John Templeton Foundation or the Center for SpaceTime and the Quantum. Claude Opus was used to improve readability.
\end{acknowledgements}

\appendix

\section{Point-singularity model of a circular treadmilling swimmer}\label{app:point_swimmer}

\begin{figure}[t]
\centering
\includegraphics[width=0.75\textwidth]{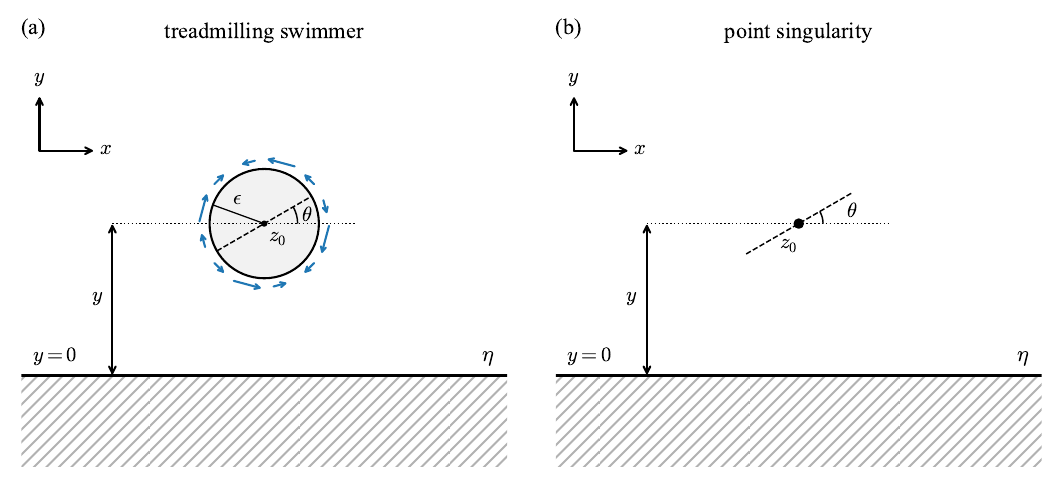}
\caption{The Crowdy--Or~\cite{crowdy2010two} circular treadmiller and its point-singularity model. (a)~A circular treadmilling swimmer of radius $\epsilon$, with centre $z_0(t)=x(t)+\ii y(t)$ and axis at angle $\theta(t)$, lies above a no-slip boundary $y=0$. Arrows on the swimmer indicate the surface velocity~\eqref{eq:slipvel}. (b)~Within a point-singularity model, the swimmer is represented by a stresslet and a source quadrupole at $z_0$.}
\label{fig:schematic_noslip}
\end{figure}

This appendix explains the origins of the point swimmer model used in this paper. It was first introduced by Crowdy and Or~\cite{crowdy2010two} who considered a model circular swimmer of radius $\epsilon$ in a two-dimensional Stokes flow, with centre at $z_0(t)=x(t)+\ii y(t)$ in the upper half plane $y>0$ near a no-slip wall and with the orientation of some distinguished diameter at angle $\theta(t)$ measured from the positive $x$ axis (Fig.~\ref{fig:schematic_noslip}). Crowdy and Or~\cite{crowdy2010two} supposed that the swimmer generates a local flow by virtue of a prescribed tangential surface velocity (caused perhaps by cilia within an envelope model) given by
\begin{equation}
U_{\mathrm{s}}(\phi,t) = 2V_{\mathrm{s}}\sin{\bigl[2\bigl(\phi-\theta(t)\bigr)\bigr]},
\label{eq:slipvel}
\end{equation}
where $\phi$ is the polar angle around the swimmer centre, measured from the positive $x$ axis, positive $U_{\mathrm{s}}$ is anticlockwise, and the constant $V_{\mathrm{s}} > 0$ sets the amplitude of the surface velocity. Because $U_{\mathrm{s}}$ contains only the second Fourier mode, the swimmer has no head or tail: $\theta$ and $\theta+\pi$ describe the same swimmer, so $\theta$ specifies the orientation of a diameter rather than a direction; the diameter $\phi=\theta,\theta+\pi$ is therefore thought of as the swimmer axis. On it the surface velocity vanishes and is directed away from it on either side, so fluid is drawn towards the swimmer along the axis and expelled along the perpendicular diameter ($\phi=\theta\pm\pi/2$). Thus $\theta=\pm\pi/2$ corresponds to this axis being normal to the boundary and, if $\theta=0$, the swimmer axis is  parallel to the boundary. A treadmilling swimmer with the tangential velocity prescribed in \eqref{eq:slipvel} does not move in an unbounded fluid, and if placed near a no-slip wall, or other type of interface or boundary, the swimmer moves only because of its interaction with it.

Crowdy and Or~\cite{crowdy2010two} showed that the flow generated by such a swimmer in free space is equivalent to a point singularity at its centre which is a superposition of a stresslet of strength $\mu=\epsilon V_{\mathrm{s}}\ee^{2\ii\theta}$ with a source quadrupole of strength $\beta=2\epsilon^2\mu$~\cite{crowdy2010two,davis2012stresslet}. On choosing the time scale so that $\epsilon V_{\mathrm{s}}=1$, this gives $\mu=\ee^{2\ii\theta}$ and $\beta=2\epsilon^2\ee^{2\ii\theta}$, which is precisely the point swimmer used throughout this study.

It is of interest to note that, after Crowdy and Or~\cite{crowdy2010two} had studied the dynamics of this point-singularity model of the swimmer near a no-slip wall it was later noticed by Crowdy~\cite{crowdy2011treadmilling} that the exact dynamical system describing such a circular treadmiller  near a wall can also be written down explicitly using the reciprocal theorem of Stokes flow combined with the known analytical solution for the dragging problem of a cylinder near a wall. Reassuringly, although, as expected, the two dynamical systems exhibit some differences, the qualitative dynamics is found to be similar.

Although this will not be used here, another point to mention is that, while replacing this finite-area treadmiller with a point swimmer with
the above stresslet-quadrupole combination is a model (that is, it is not an asymptotically exact approximation for small swimmer size in the presence of a boundary), Davis and Crowdy~\cite{davis2013matched} have carried out the next-order matched asymptotic analysis to identify the asymptotically correct equations.

\section{Derivation for the fluid--fluid interface}\label{app:interface}

This appendix derives the solution~\eqref{eq:hsol} for the fluid--fluid interface and the resulting equations of motion~\eqref{eq:ode}. The derivation proceeds in three steps. The interface conditions are first written in terms of the functions $h_\pm(z)$ and $\hat g_\pm'(z)$ introduced in~\eqref{eq:h}; these functions are then determined by Schwarz-reflection arguments, using only their singular behaviour at $z_0$; finally, the regular parts of the velocity and vorticity at $z_0$ give the swimmer velocity and rotation rate through~\eqref{eq:faxen}. To allow later extensions, the calculation is carried out for a stresslet combined with a source dipole and a source quadrupole of arbitrary strengths, and specialised to the chosen point-singularity model of a treadmilling swimmer at the end.

A useful first step is to rewrite the boundary conditions in terms of the functions $h_\pm(z)$ introduced in
(\ref{eq:h}).
In the upper ($+$) and lower ($-$) fluids the streamfunctions are written as
\begin{equation}
\psi_\pm=\Imag[\zb f_\pm(z)+g_\pm(z)]
\end{equation}
with
\begin{equation}
g_\pm(z)=-zf_\pm(z)+\hat{g}_\pm(z).
\label{eq:gdecomp}
\end{equation}
From~\eqref{eq:goursat},
\begin{equation}
u_\pm-\ii v_\pm=-\left[f_\pm(z)+\overline{f_\pm(z)}\right]-(z-\zb)f_\pm'(z)+\hat{g}_\pm'(z),
\label{eq:A_vel}
\end{equation}
so that on $y=0$, where $z=\zb$, the normal velocity vanishes provided 
\begin{equation}
\Imag[\hat{g}_\pm'(z)]=0.
\end{equation}
The slip velocity on the boundary $u_\pm\big|_{y=0}$ is then
\begin{equation}
u_\pm\big|_{y=0}=\Real[h_\pm(z)], \qquad h_\pm(z)=-2f_\pm(z)+\hat{g}_\pm'(z).
\end{equation}
The functions $f_-(z)$, $g_-(z)$ and hence $h_-(z)$ are analytic in the lower half plane. The complex form of the fluid stress tensor at the boundary, or $\sigma^{\pm}_{ij} n_j = -p_\pm n_i + 2\eta_\pm e^{\pm}_{ij} n_j$ where $e^\pm_{ij}$ is the fluid rate-of-strain tensor, is
\begin{equation}
-2 \eta_\pm \ii \frac{\dd H_\pm(z, \zb)}{\dd x}, \qquad H_\pm(z,\zb) = f_\pm(z) + z \overline{f_\pm'(z)} + \overline{g_\pm'(z)}.
\label{eq:stressform}
\end{equation}
On use of \eqref{eq:gdecomp} it can be shown that
\begin{equation}
\Imag[H_\pm(z, \zb)] = - \Imag[h_\pm(z)] \qquad \text{on}~y=0.
\end{equation}

For the two-fluid problem with a flat interface between them at $y=0$ the conditions are that the tangential slip velocities and tangential stresses are continuous, namely,
\begin{equation}
u_+\big|_{y=0} = u_-\big|_{y=0}, \qquad t_i \sigma^+_{ij}n_j = t_i \sigma^-_{ij}n_j.
\end{equation}
The boundary conditions are therefore
\begin{equation}
\Real[h_+(z)]=\Real[h_-(z)],\qquad \Imag[h_+(z)]=\sigma\,\Imag[h_-(z)] \qquad\text{on}~y=0,
\label{eq:A_bc}
\end{equation}
where the viscosity ratio $\sigma$ is given by \eqref{eq:sigma}. The second of these is an integral with respect to $x$ of the tangential stress condition after use of the fact that
\begin{equation}
t_i \sigma^\pm_{ij}n_j = \Real\left[-2 \eta_\pm \ii \frac{\dd H_\pm(z,\zb)}{\dd x}\right] = -\Imag\left[2 \eta_\pm \frac{\dd h_\pm(z)}{\dd x} \right].
\label{eq:traction_h}
\end{equation}

\subsection{Determination of $\hat{g}_\pm(z)$ and $h_\pm(z)$}

The reason for introducing the two functions $\hat g_\pm(z)$ and $h_\pm(z)$ is that both can be shown to satisfy simple Schwarz-type boundary value problems in a half-plane. Such problems can be solved by Schwarz reflection, as will now be demonstrated.

First, it is necessary to ascertain the singularity structure of each function. 
Even though it is not used in the swimmer model, a source dipole is also included since it changes little in the
structure of the analysis.
A stresslet of strength $\mu$, a source dipole of strength $\gamma$ and a source quadrupole of strength $\beta$ at $z_0$ correspond to a local singularity structure given by
\begin{equation}
f_+(z)=\frac{\mu}{z-z_0}+(\text{analytic}),\qquad
g_+'(z)=\frac{\mu\zob+\gamma}{(z-z_0)^2}+\frac{\beta}{(z-z_0)^3}+(\text{analytic}),
\end{equation}
where simple poles in $g_+'(z)$ are excluded to avoid a net source or rotlet. Since $\hat{g}_+'(z)=g_+'(z)+f_+(z)+zf_+'(z)$, it follows that near $z_0$
\begin{equation}
\hat{g}_+'(z)=\frac{B}{(z-z_0)^2}+\frac{C}{(z-z_0)^3}+(\text{analytic}),\qquad B=\mu(\zob-z_0)+\gamma,\quad C=\beta .
\end{equation}
With the singularities of $\hat{g}_+'(z)$ in the upper half plane determined, and since $\hat{g}_+'(z)$ is real on $y=0$, 
Schwarz reflection gives
\begin{equation}
\hat{g}_+'(z)=\frac{B}{(z-z_0)^2}+\frac{C}{(z-z_0)^3}+\frac{\overline{B}}{(z-\zob)^2}+\frac{\overline{C}}{(z-\zob)^3},
\label{eq:A_ghat}
\end{equation}
and therefore $h_+(z)=-2f_+(z)+\hat{g}_+'(z)$ has the singular part 
\begin{equation}
S(z)=\frac{A}{z-z_0}+\frac{B}{(z-z_0)^2}+\frac{C}{(z-z_0)^3}
\end{equation}
with $A=-2\mu$, as in \eqref{eq:hlocal}. Note that $\gamma=0$ for the treadmilling swimmer treated in the paper.

With $\hat{g}'(z)$ determined, it only remains to determine $h_+(z)$. Let 
\begin{equation}
\overline{S}(z)=\overline{S(\zb)}=\frac{\overline{A}}{z-\zob}+\frac{\overline{B}}{(z-\zob)^2}+\frac{\overline{C}}{(z-\zob)^3},
\end{equation}
which is analytic in the upper half plane, and similarly $\overline{h_-}(z)=\overline{h_-(\zb)}$. On $y=0$ the real parts of a function and of its reflection coincide, so the first condition in \eqref{eq:A_bc} can be written $\Real[h_+(z)-S(z)]=\Real[\overline{h_-}(z)-\overline{S}(z)]$. Both bracketed functions are analytic in the upper half plane and decay at infinity, so they are equal:
\begin{equation}
h_+(z)-S(z)=\overline{h_-}(z)-\overline{S}(z).
\label{eq:A_cont1}
\end{equation}
In the same way, since imaginary parts change sign under reflection, the second condition gives
\begin{equation}
h_+(z)-S(z)=-\sigma\overline{h_-}(z)+\overline{S}(z).
\label{eq:A_cont2}
\end{equation}
Eliminating $\overline{h_-}(z)$ between \eqref{eq:A_cont1} and \eqref{eq:A_cont2} gives $(1+\sigma)[h_+(z)-S(z)]=(1-\sigma)\overline{S}(z)$, which is \eqref{eq:hsol}. 

Finally, from $f_+(z)=[-h_+(z)+\hat{g}_+'(z)]/2$ with \eqref{eq:hsol} and \eqref{eq:A_ghat},
\begin{equation}
f_+(z)=\frac{\mu}{z-z_0}+\frac{1-\sigma}{1+\sigma}\frac{\overline{\mu}}{z-\zob}+\frac{\sigma}{1+\sigma}\left[\frac{\overline{B}}{(z-\zob)^2}+\frac{\overline{C}}{(z-\zob)^3}\right],
\label{eq:A_f}
\end{equation}
with a similar expression for $f_-(z)$. The complex velocity field of the upper fluid is now given by
\begin{equation}
u_+ - \ii v_+ =- \left[f_+(z) + \overline{f_+(z)}\right] - (z-\zb)f_+'(z) + \hat{g}_+'(z).
\label{eq:uvexp}
\end{equation}

It should be noted that a similar construction applied to a Stokeslet gives the two-dimensional analogue of the image system of Aderogba and Blake~\cite{aderogba1978action} for a point force near the interface between two immiscible liquids, which reduces to Blake's image system for a no-slip wall~\cite{blake1971note} as $\sigma\to\infty$. 

\subsection{Dynamical system}

With the flow field determined, the dynamics of the swimmer can be found. The terms in the velocity field~\eqref{eq:uvexp} that are singular at $z_0$ are those of the swimmer in unbounded fluid; the remaining terms form the image system. Expanding \eqref{eq:uvexp} and $4f_+'(z)$ about $z_0$ and retaining the regular parts allows determination of the swimmer speed $(U,V)$ and its angular velocity $\omega_0/2$ from
\begin{align}
U-\ii V&=\frac{\mu+(1+2\sigma)\overline{\mu}}{(1+\sigma)(z_0-\zob)}+\frac{-\sigma\gamma+(1+2\sigma)\overline{\gamma}}{(1+\sigma)(z_0-\zob)^2}+\frac{\sigma\beta+(1+3\sigma)\overline{\beta}}{(1+\sigma)(z_0-\zob)^3},\label{eq:A_UV}\\
\frac{p_0}{\eta_+}-\ii\omega_0&=-4\left[\frac{\overline{\mu}}{(z_0-\zob)^2}+\frac{2\sigma}{1+\sigma}\frac{\overline{\gamma}}{(z_0-\zob)^3}+\frac{3\sigma}{1+\sigma}\frac{\overline{\beta}}{(z_0-\zob)^4}\right].\label{eq:A_omega}
\end{align}
Substituting $\mu=\ee^{2\ii\theta}$, $\gamma=0$, $\beta=2\epsilon^2\mu$ and $z_0-\zob=2\ii y$ into \eqref{eq:A_UV}--\eqref{eq:A_omega}, and using \eqref{eq:faxen}, yields the equations of motion~\eqref{eq:ode}. In the limit $\sigma\to\infty$ these reduce to the no-slip-wall results of \cite{crowdy2010two}.

\section{Derivation for the partial-slip wall}\label{app:slip}

This appendix derives the equations of motion~\eqref{eq:ode_slip}--\eqref{eq:Fthetaslip} for the partial-slip wall. The construction follows that of Appendix~\ref{app:interface}, with one difference: the Navier-slip condition becomes a first-order linear differential equation for $h(z)$ on the wall, rather than an algebraic relation between $h_+(z)$ and $h_-(z)$. This equation is solved with an integrating factor, which leads to exponential integrals and to an image system consisting of a point image together with a continuous line distribution of singularities. The swimmer velocity and rotation rate then follow, as before, from the regular parts of the flow at $z_0$.

For the partial-slip wall there is fluid only in $y>0$, and we drop the subscript $+$. The representation~\eqref{eq:A_vel}, the reflection formula~\eqref{eq:A_ghat} and the singular part $S(z)$ of $h(z)$ are unchanged. The Navier-slip condition on the wall is
\begin{equation}
u_i t_i = {\lambda \over \eta} t_i \sigma_{ij} n_j.
\end{equation}
Since $u_i t_i = u\big|_{y=0}$ and since, from \eqref{eq:traction_h}, 
\begin{equation}
t_i \sigma_{ij} n_j = - 2 \eta \Imag[h'(z)] 
\end{equation}
then the Navier condition~\eqref{eq:navier} takes the form
\begin{equation}
\Real\left[h(z)+\ii\kappa h'(z)\right]=0\qquad\text{on } y=0,\qquad \kappa=-2\lambda .
\label{eq:B_bc}
\end{equation}

\subsection{Determination of $h(z)$}\label{app:slip_image}

The function
\begin{equation}
R(z)=\left[S(z)+\ii\kappa S'(z)\right]-\left[\overline{S}(z)-\ii\kappa\overline{S}'(z)\right]
\end{equation}
has purely imaginary values on $y=0$ and the same singular part at $z_0$ as $h(z)+\ii\kappa h'(z)$. Hence $h(z)+\ii\kappa h'(z)-R(z)$ is analytic in the upper half plane, decays at infinity and has zero real part on $y=0$, so it vanishes identically. This implies, by analytic continuation off $y=0$, that $h(z)$ satisfies the first-order linear differential equation
\begin{equation}
h'(z)+\frac{h(z)}{\ii\kappa}=\frac{R(z)}{\ii\kappa}.
\label{eq:B_ode}
\end{equation}
Equation~\eqref{eq:B_ode} is solved with the integrating factor $\ee^{z/(\ii\kappa)}$, choosing the constant of integration so that $h(z)$ decays at infinity in the upper half plane. Each term of $R(z)$ is of the form $D/(z-z_*)^p$ with $D$ constant and $z_*=z_0$ or $\zob$, and the corresponding integral is
\begin{equation}
\int^z\frac{\ee^{z'/(\ii\kappa)}}{\ii\kappa}\frac{D}{(z'-z_*)^p}\,\dd z'=-\frac{D\,\ee^{z_*/(\ii\kappa)}}{\ii\kappa\,(z-z_*)^{p-1}}\,\EE_p\!\left(-\frac{z-z_*}{\ii\kappa}\right),
\end{equation}
where $\EE_p(w)=\int_1^\infty \ee^{-ws}s^{-p}\,\dd s$ is the generalised exponential integral of order $p$~\cite{abramowitz1965handbook,oldham2009atlas,olver2010nist}. Since $R(z)=T(z)-\overline{T}(z)$ with $T(z)=S(z)+\ii\kappa S'(z)$, we may write
\begin{equation}
R(z)=\sum_{p=1}^{4}\left[\frac{a_p}{(z-z_0)^p}-\frac{\overline{a_p}}{(z-\zob)^p}\right],
\end{equation}
with $a_1=A$, $a_2=B-\ii\kappa A$, $a_3=C-2\ii\kappa B$ and $a_4=-3\ii\kappa C$, and the solution is
\begin{equation}
h(z)=-\frac{1}{\ii\kappa}\sum_{p=1}^{4}\left[\frac{a_p\,\ee^{w_0}\EE_p(w_0)}{(z-z_0)^{p-1}}-\frac{\overline{a_p}\,\ee^{\tilde{w}_0}\EE_p(\tilde{w}_0)}{(z-\zob)^{p-1}}\right],
\label{eq:B_h}
\end{equation}
where $w_0=-(z-z_0)/(\ii\kappa)$ and $\tilde{w}_0=-(z-\zob)/(\ii\kappa)$.

The structure of the image system becomes clear if the swimmer's own singularity is separated out, $h(z)=S(z)+h_{\mathrm{img}}(z)$. Since $R(z)=T(z)-\overline{T}(z)$ with $T(z)=S(z)+\ii\kappa S'(z)$, \eqref{eq:B_ode} gives
\begin{equation}
h_{\mathrm{img}}'(z)+\frac{h_{\mathrm{img}}(z)}{\ii\kappa}=-\frac{\overline{T}(z)}{\ii\kappa},
\end{equation}
whose solution decaying at infinity is
\begin{equation}
h_{\mathrm{img}}(z)=-\frac{1}{2\lambda}\int_0^\infty \ee^{-\ell/(2\lambda)}\,\overline{T}(z+\ii\ell)\,\dd\ell
=\overline{S}(z)-\frac{1}{\lambda}\int_0^\infty \ee^{-\ell/(2\lambda)}\,\overline{S}(z+\ii\ell)\,\dd\ell ,
\label{eq:B_image}
\end{equation}
where the second form follows on integrating the $\overline{S}'(z)$ term by parts. The first term in \eqref{eq:B_image} is the image for a perfectly slipping wall, a single reflected singularity at the mirror point $\zob$. In the second, $\overline{S}(z+\ii\ell)$ has its poles at $z=\zob-\ii\ell$, a distance $\ell$ below the mirror point, so the integral is a continuous distribution of the same reflected singularities along the ray extending downwards from $\zob$, with weight decaying exponentially over the depth $2\lambda$. As $\lambda\to0$ the distribution collapses onto $\zob$ and \eqref{eq:B_image} reduces to $-\overline{S}(z)$, the no-slip image; as $\lambda\to\infty$ it recedes to infinity, leaving $\overline{S}(z)$, the free-surface image. A similar structure, a primary image together with singularities distributed along a line with weights decaying exponentially over the slip length, 
was found in three dimensions~\cite{lauga2005brownian} and used for swimmers in \cite{lopez2014dynamics}.
To the best of the authors' knowledge, the treatment above appears to be the first time that the analogous
problem has been treated in the two-dimensional case. It is interesting to note that the approach using
complex analysis methods reduces the problem to a linear ordinary differential equation.

\subsection{Dynamical system}

The Goursat function $f(z)=[-h(z)+\hat g'(z)]/2$ and the regular parts of the velocity and vorticity at $z_0$ then follow as in Appendix~\ref{app:interface}. The algebra is lengthy but is simplified by the recurrence 
relation
$n\EE_{n+1}(w)=\ee^{-w}-w\EE_n(w)$ and the derivative relation $\EE_n'(w)=-\EE_{n-1}(w)$, which reduce all exponential integrals to $\EE_1(w)$~\cite{abramowitz1965handbook,oldham2009atlas,olver2010nist}. At the swimmer centre the argument of the image terms is $-(z_0-\zob)/(\ii\kappa)=y/\lambda$, which is real and positive. For a stresslet $\mu$ with a source dipole $\gamma$ and a source quadrupole $\beta$ at $z_0$, so that $A=-2\mu$, $B=\mu(\zob-z_0)+\gamma$ and $C=\beta$, the result is
\begin{align}
U-\ii V &= \frac{\ee^{-\frac{z_0-\zob}{\ii\kappa}}}{2\kappa^4} \EE_1\left(-\frac{z_0-\zob}{\ii\kappa}\right) \{ -2\ii\kappa\overline{\mu} (z_0-\zob)^2 + [2\kappa^2(\mu-3\overline{\mu})-2\ii\kappa\overline{\gamma}-\overline{\beta}] (z_0-\zob) + [-4\ii\kappa^3(\mu-\overline{\mu})-2\kappa^2(\gamma+\overline{\gamma})-\ii\kappa(\beta-\overline{\beta})] \} \nonumber \\
&\quad + \frac{\overline{\mu}}{\kappa^2} (z_0-\zob) + \frac{2\ii\kappa^2(\mu-2\overline{\mu})+2\kappa\overline{\gamma}-\ii\overline{\beta}}{2\kappa^3} + \frac{2\kappa^2(\mu+\overline{\mu})-2\ii\kappa\gamma+\beta}{2\kappa^2} \frac{1}{z_0-\zob} + \frac{2\kappa\overline{\gamma}+\ii(\beta+\overline{\beta})}{2\kappa} \frac{1}{(z_0-\zob)^2} + \frac{\overline{\beta}}{(z_0-\zob)^3} ,
\label{eq:B_UV} \\
\frac{p_0}{\eta} - \ii\omega_0 &= \frac{2\ee^{-\frac{z_0-\zob}{\ii\kappa}}}{\kappa^4} \EE_1\left(-\frac{z_0-\zob}{\ii\kappa}\right) [2\ii\kappa\overline{\mu}(z_0-\zob) + 4\kappa^2\overline{\mu} + 2\ii\kappa\overline{\gamma} + \overline{\beta}] \nonumber \\
&\quad + 2 \left\{ -\frac{2\overline{\mu}}{\kappa^2} + \frac{2\ii\kappa^2\overline{\mu}-2\kappa\overline{\gamma}+\ii\overline{\beta}}{\kappa^3} \frac{1}{z_0-\zob} + \frac{-2\kappa^2\overline{\mu}-2\ii\kappa\overline{\gamma}-\overline{\beta}}{\kappa^2} \frac{1}{(z_0-\zob)^2} - \frac{2\ii\overline{\beta}}{\kappa} \frac{1}{(z_0-\zob)^3} \right\} .
\label{eq:B_omega}
\end{align}
Substituting $\kappa=-2\lambda$, $z_0-\zob=2\ii y$ and the treadmilling-swimmer strengths $\mu=\ee^{2\ii\theta}$, $\gamma=0$, $\beta=2\epsilon^2\mu$ yields~\eqref{eq:ode_slip}--\eqref{eq:Fthetaslip}. As a check, the small-argument and large-argument behaviour of $\EE_1(w)$ recovers the limits $\lambda\to\infty$ and $\lambda\to0$ (Appendix~\ref{app:asym}).

\section{Stresslet functions and asymptotic expansions}\label{app:asym}

This appendix collects the expansions used in Section~\ref{sec:slip}. Appendix~\ref{app:stresslet_slip} gives the stresslet functions $s_x(\tilde{y})$, $s_y(\tilde{y})$ and $s_\theta(\tilde{y})$ of \eqref{eq:stresslet_slip} in closed form, together with their limiting behaviour. Appendices~\ref{app:wall_asym} and~\ref{app:free_asym} expand the slip-wall equations~\eqref{eq:ode_slip} for small and large slip length and compare them with the interface equations~\eqref{eq:ode} in the corresponding limits $\sigma\to\infty$ and $\sigma\to0$, showing that the leading terms coincide and how the corrections differ. Throughout, $\gamma_{\mathrm E}\approx0.5772$ denotes Euler's constant.

\subsection{Pure stresslet near a partial-slip wall}\label{app:stresslet_slip}

Setting $\tilde\epsilon=0$ in \eqref{eq:Fxslip}--\eqref{eq:Fthetaslip} and comparing \eqref{eq:ode_slip} with \eqref{eq:stresslet_slip} gives $s_x(\tilde{y})=\tilde{y}\mathcal{F}_x(\tilde{y};0)/2$, $s_y(\tilde{y})=\tilde{y}\mathcal{F}_y(\tilde{y};0)/2$ and $s_\theta(\tilde{y})=\tilde{y}^2\mathcal{F}_\theta(\tilde{y};0)$, that is,
\begin{align}
s_x(\tilde{y})&=\frac{1}{2}\left[\tilde{y}(\tilde{y}+2)^2\,\ee^{\tilde{y}}\EE_1(\tilde y)-\tilde{y}^2-3\tilde{y}\right],\\
s_y(\tilde{y})&=\frac{1}{2}\left[\tilde{y}^2(\tilde{y}+2)\,\ee^{\tilde{y}}\EE_1(\tilde y)-\tilde{y}^2-\tilde{y}+2\right],\\
s_\theta(\tilde{y})&=\tilde{y}^2(\tilde{y}+2)\,\ee^{\tilde{y}}\EE_1(\tilde{y})-\tilde{y}^2-\tilde{y}+1 .
\end{align}
The relation $s_\theta(\tilde{y})=2s_y(\tilde{y})-1$ follows directly, and is the case $\epsilon=0$ of \eqref{eq:nullcline_identity}; it holds trivially for the interface. Numerical evaluation over $10^{-5}\le\tilde{y}\le10^4$ shows that $s_x(\tilde{y})$ increases monotonically from $0$ to $1$, and that $s_y(\tilde{y})$ has a single minimum $s_y\approx0.890$ at $\tilde y\approx1.64$, where $s_\theta\approx0.780$. For $\tilde{y}\gg1$, $s_x(\tilde{y})=1-\tilde y^{-1}+4\tilde{y}^{-2}+O(\tilde{y}^{-3})$, $s_y(\tilde{y})=1-\tilde{y}^{-1}+6\tilde{y}^{-2}+O(\tilde{y}^{-3})$ and $s_\theta(\tilde{y})=1-2\tilde{y}^{-1}+12\tilde{y}^{-2}+O(\tilde{y}^{-3})$, and for $\tilde{y}\ll1$,
\begin{equation}
s_x(\tilde{y})=\left(2\log\frac{1}{\tilde{y}}-2\gamma_{\mathrm E}-\frac32\right)\tilde{y}+O(\tilde{y}^2\log\tilde{y}),\qquad
s_y(\tilde{y})=1-\frac{\tilde{y}}{2}+O(\tilde{y}^2\log\tilde{y}),
\end{equation}
consistent with \eqref{eq:U_large_slip}--\eqref{eq:omega_large_slip} at $\epsilon=0$, with $s_\theta(\tilde{y})=2s_y(\tilde{y})-1=1-\tilde{y}+O(\tilde{y}^2\log\tilde{y})$. The approach of $s_x(\tilde{y})$ to its free-surface value is therefore logarithmic, whereas for the interface $s_x\sim\sigma$ as $\sigma\to0$.

\subsection{Small slip length and large viscosity ratio}\label{app:wall_asym}

Expanding \eqref{eq:ode_slip} for $\lambda\ll y$, using the large-argument expansion of $\EE_1(w)$~\cite{abramowitz1965handbook,oldham2009atlas,olver2010nist}, gives
\begin{align}
\dot{x} &= \frac{\sin{2\theta}}{y} \left[ -\left(1-\frac{\epsilon^2}{2y^2}\right) + \left(1-\frac{3\epsilon^2}{2y^2}\right) \frac{\lambda}{y} - 4\left(1-\frac{9\epsilon^2}{4y^2}\right) \frac{\lambda^2}{y^2} + O\left(\frac{\lambda^3}{y^3}\right) \right] , \\
\dot{y} &= \frac{\cos{2\theta}}{y} \left[ \left(1-\frac{\epsilon^2}{y^2}\right) - \left(1-\frac{3\epsilon^2}{y^2}\right) \frac{\lambda}{y} + 6\left(1-\frac{5\epsilon^2}{2y^2}\right) \frac{\lambda^2}{y^2} + O\left(\frac{\lambda^3}{y^3}\right) \right] , \\
\dot{\theta} &= \frac{\sin{2\theta}}{2y^2} \left[ \left(1-\frac{3\epsilon^2}{2y^2}\right) - 2\left(1-\frac{3\epsilon^2}{y^2}\right) \frac{\lambda}{y} + 12\left(1-\frac{5\epsilon^2}{2y^2}\right) \frac{\lambda^2}{y^2} + O\left(\frac{\lambda^3}{y^3}\right) \right] .
\end{align}

The corresponding expansion of \eqref{eq:ode} for $\sigma\gg1$ is
\begin{align}
\dot{x} &= \frac{\sin{2\theta}}{y} \left[ -\left(1-\frac{\epsilon^2}{2y^2}\right) + \left(1-\frac{\epsilon^2}{4y^2}\right) \frac{1}{\sigma} - \left(1-\frac{\epsilon^2}{4y^2}\right) \frac{1}{\sigma^2} + O(\sigma^{-3}) \right] , \\
\dot{y} &= \frac{\cos{2\theta}}{y} \left[ \left(1-\frac{\epsilon^2}{y^2}\right) + \frac{3\epsilon^2}{4y^2} \frac{1}{\sigma} - \frac{3\epsilon^2}{4y^2} \frac{1}{\sigma^2} + O(\sigma^{-3}) \right] , \\
\dot{\theta} &= \frac{\sin{2\theta}}{2y^2} \left[ \left(1-\frac{3\epsilon^2}{2y^2}\right) + \frac{3\epsilon^2}{2y^2} \frac{1}{\sigma} - \frac{3\epsilon^2}{2y^2} \frac{1}{\sigma^2} + O(\sigma^{-3}) \right] .
\end{align}

The leading terms coincide, but the first corrections differ in structure. The $O(\lambda/y)$ terms of the slip-wall expansions are exactly $\lambda\,\partial_y$ of the leading terms, as can be checked directly, so that, at first order, the slip wall acts as a no-slip wall displaced to $y=-\lambda$; this equivalence fails at $O(\lambda^2)$. The three characteristic heights therefore all decrease together,
\begin{equation}
y_x\approx\frac{\epsilon}{\sqrt{2}}-\lambda,\qquad y_y\approx\epsilon-\lambda,\qquad y_\theta\approx\sqrt{\frac{3}{2}}\,\epsilon-\lambda .
\label{eq:heights_small_slip}
\end{equation}
For the interface, by contrast, the $1/\sigma$ corrections to the normal velocity and rotation rate involve only the quadrupole terms, consistent with the $\sigma$-independence of these rates for the stresslet alone (Section~\ref{sec:interface}), whereas slip already modifies the stresslet contribution at $O(\lambda/y)$. The heights therefore move in opposite directions,
\begin{equation}
y_x\approx\frac{\epsilon}{\sqrt{2}}\left(1+\frac{1}{4\sigma}\right),\qquad
y_y\approx\epsilon\left(1-\frac{3}{8\sigma}\right),\qquad
y_\theta\approx\sqrt{\frac{3}{2}}\,\epsilon\left(1-\frac{1}{2\sigma}\right).
\label{eq:heights_large_sigma}
\end{equation}

\subsection{Large slip length and small viscosity ratio}\label{app:free_asym}

Expanding~\eqref{eq:ode_slip} for $\lambda\gg y$, using the small-argument expansion $\EE_1(w)=-\gamma_{\mathrm E}-\log w+w+O(w^2)$~\cite{abramowitz1965handbook,oldham2009atlas,olver2010nist}, gives
\begin{align}
\dot{x} &= \frac{\sin{2\theta}}{y} \left[ \frac{\epsilon^2}{4y^2} + \left(2\log{\frac{y}{\lambda}}+2\gamma_{\mathrm E}+\frac{3}{2}\right) \frac{y}{\lambda} + \left(4\log{\frac{y}{\lambda}}+4\gamma_{\mathrm E}-\frac{3}{2}+\frac{\epsilon^2}{8y^2}\right) \frac{y^2}{\lambda^2} + o\left(\frac{y^2}{\lambda^2}\right) \right] , \label{eq:U_large_slip} \\
\dot{y} &= \frac{\cos{2\theta}}{y} \left[ \left(1-\frac{\epsilon^2}{4y^2}\right) - \frac{1}{2}\left(1+\frac{\epsilon^2}{2y^2}\right) \frac{y}{\lambda} + \left(-\log{\frac{y}{\lambda}}-\gamma_{\mathrm E}-\frac{1}{2}+\frac{\epsilon^2}{8y^2}\right) \frac{y^2}{\lambda^2} + o\left(\frac{y^2}{\lambda^2}\right) \right] , \label{eq:V_large_slip} \\
\dot{\theta} &= \frac{\sin{2\theta}}{2y^2} \left[ 1 - \left(1+\frac{\epsilon^2}{2y^2}\right) \frac{y}{\lambda} + \left(-2\log{\frac{y}{\lambda}}-2\gamma_{\mathrm E}-1+\frac{\epsilon^2}{4y^2}\right) \frac{y^2}{\lambda^2} + o\left(\frac{y^2}{\lambda^2}\right) \right] . \label{eq:omega_large_slip}
\end{align}

The corresponding expansion of \eqref{eq:ode} for $\sigma\ll1$ is
\begin{align}
\dot{x} &= \frac{\sin{2\theta}}{y} \left[ \frac{\epsilon^2}{4y^2} - \left(1-\frac{\epsilon^2}{4y^2}\right) \sigma + \left(1-\frac{\epsilon^2}{4y^2}\right) \sigma^2 + O(\sigma^3) \right] , \\
\dot{y} &= \frac{\cos{2\theta}}{y} \left[ \left(1-\frac{\epsilon^2}{4y^2}\right) - \frac{3\epsilon^2}{4y^2} \sigma + \frac{3\epsilon^2}{4y^2} \sigma^2 + O(\sigma^3) \right] , \\
\dot{\theta} &= \frac{\sin{2\theta}}{2y^2} \left[ 1 - \frac{3\epsilon^2}{2y^2} \sigma + \frac{3\epsilon^2}{2y^2} \sigma^2 + O(\sigma^3) \right] .
\end{align}

Again the leading terms coincide and the corrections differ; in particular, the slip-wall corrections contain $\log(\lambda/y)$.

\section{Equilibria, stability and first integrals}\label{app:stability}

This appendix treats the interface and the slip wall together. Appendix~\ref{app:nullcline} shows that both systems share the same angular structure and an identity, independent of the boundary type, that fixes the transition height $y^*=\epsilon/\sqrt2$. Appendix~\ref{app:stability_eq} classifies the equilibria and their stability, and Appendix~\ref{app:firstintegral} derives the first integrals used in Sections~\ref{sec:interface} and~\ref{sec:slip}.

\subsection{Common structure of the equations}\label{app:nullcline}

Both the interface equations~\eqref{eq:ydot}--\eqref{eq:thdot} and the slip-wall equations~\eqref{eq:ode_slip} have the form
\begin{equation}
\dot y=\cos2\theta\,a(y),\qquad \dot\theta=\sin2\theta\,b(y),
\label{eq:ab_form}
\end{equation}
with
\begin{equation}
a(y)=\frac{1}{y}\left(1-\frac{y_y^2}{y^2}\right),\qquad b(y)=\frac{1}{2y^2}\left(1-\frac{y_\theta^2}{y^2}\right)
\end{equation}
for the interface, and 
\begin{equation}
    a(y)=\frac{\mathcal{F}_y(\tilde y;\tilde\epsilon)}{2\lambda},\qquad b(y)=\frac{\mathcal{F}_\theta(\tilde y;\tilde\epsilon)}{2\lambda^2}
\end{equation}
for the slip wall. We denote by $y_a$ and $y_b$ the roots of $a(y)$ and $b(y)$, that is $y_y$ and $y_\theta$ for both boundaries. In both cases $a(y)$ and $b(y)$ are negative below these simple roots and positive above them: for the interface this is immediate, and for the slip wall $\mathcal{F}_y(\tilde{y};\tilde{\epsilon})$ and $\mathcal{F}_\theta(\tilde{y};\tilde{\epsilon})$ tend to $-\infty$ as $\tilde{y}\to0^+$ and are positive as $\tilde{y}\to\infty$, with a single root each (Section~\ref{sec:slip_equilibria}).

For both boundaries, and for all $\sigma$ and $\lambda$, the coefficients satisfy
\begin{equation}
a(y)-y\,b(y)=\frac{1}{2y}\left(1-\frac{\epsilon^2}{2y^2}\right).
\label{eq:nullcline_identity}
\end{equation}
For the interface this follows from $a(y)-y\,b(y)=(2y)^{-1}\left[1-(2y_y^2-y_\theta^2)/y^2\right]$ and $2y_y^2-y_\theta^2=\epsilon^2/2$, by \eqref{eq:heights}; all dependence on $\sigma$ cancels. For the slip wall, the exponential-integral terms in \eqref{eq:Fyslip} and \eqref{eq:Fthetaslip} cancel in the combination
\begin{equation}
\mathcal{F}_y(\tilde{y};\tilde{\epsilon})-\tilde y\,\mathcal{F}_\theta(\tilde{y};\tilde{\epsilon})=\frac{1}{\tilde y}\left(1-\frac{\tilde\epsilon^2}{2\tilde y^2}\right),
\end{equation}
and since $a(y)-y\,b(y)=[\mathcal{F}_y(\tilde{y};\tilde{\epsilon})-\tilde{y}\,\mathcal{F}_\theta(\tilde{y};\tilde{\epsilon})]/(2\lambda)$, with $\lambda\tilde{y}=y$ and $\tilde\epsilon/\tilde y=\epsilon/y$, this gives \eqref{eq:nullcline_identity}. For the pure stresslet ($\epsilon=0$) the right-hand side of \eqref{eq:nullcline_identity} is $1/(2y)>0$, so the quadrupole is essential for what follows.

Evaluating \eqref{eq:nullcline_identity} at the two roots gives
\begin{equation}
b(y_a)=-\frac{1}{2y_a^2}\left(1-\frac{y^{*2}}{y_a^2}\right),\qquad
a(y_b)=\frac{1}{2y_b}\left(1-\frac{y^{*2}}{y_b^2}\right),\qquad y^*=\frac{\epsilon}{\sqrt2}.
\label{eq:ab_roots}
\end{equation}
If $y_a<y_b$, then, since $b(y)$ is negative below its root and $a(y)$ positive above its root, $b(y_a)<0$ and $a(y_b)>0$, so by \eqref{eq:ab_roots} both roots lie above $y^*$; if $y_a>y_b$, the signs are reversed and both lie below $y^*$. The two roots therefore lie on the same side of $y^*$, and can coincide only at $y^*$. For both boundaries the transition occurs at this height, and the threshold follows from a single condition: $y_\theta=\epsilon/\sqrt2$ gives $\sigma_c=1/2$, and $\tilde\epsilon_c$ is the root of $\mathcal F_\theta(\tilde\epsilon/\sqrt2;\tilde\epsilon)=0$.

For the slip wall, \eqref{eq:nullcline_identity} also shows that $\mathcal F_\theta(\tilde y^*;\tilde\epsilon)=0$ implies $\mathcal F_y(\tilde y^*;\tilde\epsilon)=0$, so the critical value is the root of the single equation $\mathcal F_\theta(\tilde\epsilon/\sqrt2;\tilde\epsilon)=0$. Such a root exists, since $y_\theta\to\sqrt{3/2}\,\epsilon>y^*$ as $\lambda\to0$ and $y_\theta\to0<y^*$ as $\lambda\to\infty$; numerically it is unique, with $\tilde\epsilon_c=1.652776$, i.e.\ $\lambda_c=0.605043\,\epsilon$.

\subsection{Equilibria and their stability}\label{app:stability_eq}

The equilibria of \eqref{eq:ab_form} are $(y_a,0)$, $(y_a,\pi/2)$ and $(y_b,\pm\pi/4)$, modulo $\pi$ in $\theta$. At $(y_a,\pi/2)$ the Jacobian $J(y,\theta)$ is diagonal, with eigenvalues $-a'(y_a)<0$ and $-2b(y_a)$, so by \eqref{eq:ab_roots} this point is a stable node if $y_a<y^*$ and a saddle if $y_a>y^*$; at $(y_a,0)$ the signs are reversed, giving an unstable node or a saddle, respectively. At $(y_b,\pi/4)$ the diagonal entries vanish, so $\operatorname{tr}J(y_b,\pi/4)=0$ and $\det J(y_b,\pi/4)=2a(y_b)\,b'(y_b)$ with $b'(y_b)>0$; by \eqref{eq:ab_roots} this point is a linear centre if $y_b>y^*$ and a saddle if $y_b<y^*$. Since $y_a$ and $y_b$ lie on the same side of $y^*$, all equilibria change type together. For the interface, $y_y<y^*$, equivalently $y_\theta<y^*$, holds exactly for $\sigma<1/2$; for the slip wall it holds for $\lambda>\lambda_c$. This establishes the classifications of Sections~\ref{sec:equilibria} and~\ref{sec:slip_equilibria}.

For the interface the eigenvalues are explicit. At $(y_y,0)$, $J(y_y,0)=\operatorname{diag}\bigl(2/y_y^2,\ (1-y_\theta^2/y_y^2)/y_y^2\bigr)$, with $J(y_y,0)=-J(y_y,\pi/2)$, and at $(y_\theta,\pi/4)$, $\partial\dot y/\partial\theta=-(2/y_\theta)(1-y_y^2/y_\theta^2)$ and $\partial\dot\theta/\partial y=1/y_\theta^3$. In terms of $\sigma$ and $\epsilon$,
\begin{equation}
\det J(y_y,0)=\frac{32(1+\sigma)^2(1-2\sigma)}{\epsilon^4(1+4\sigma)^3},\qquad
\det J(y_\theta,\pi/4)=-\frac{4(1+\sigma)^2(1-2\sigma)}{27\epsilon^4\sigma^3}.
\end{equation}

At the transition, $y_a=y_b=y^*$ and $a(y^*)=b(y^*)=0$, so every point $(y^*,\theta)$ is an equilibrium. There the $\theta$-derivatives vanish, and
\begin{equation}
J=\begin{pmatrix}a'(y^*)\cos2\theta & 0\\ b'(y^*)\sin2\theta & 0\end{pmatrix},
\end{equation}
with eigenvalue $0$ along the line and $a'(y^*)\cos2\theta$ transverse to it. For the interface $a'(y^*)=2/y^{*2}=4/\epsilon^2$, so the line is normally attracting where $\cos2\theta<0$.

\subsection{First integrals}\label{app:firstintegral}

For any system of the form~\eqref{eq:ab_form}, $d(\log\sin2\theta)/dy=2b(y)/a(y)$, so
\begin{equation}
\sin2\theta\,\exp\left[-2\int\frac{b(y)}{a(y)}\,\dd y\right]
\end{equation}
is conserved. For the interface,
\begin{equation}
\frac{\dd}{\dd y}\log \sin2\theta=\frac{y^2-y_\theta^2}{y\,(y^2-y_y^2)}=\frac{y_\theta^2}{y_y^2}\frac{1}{y}+\left(1-\frac{y_\theta^2}{y_y^2}\right)\frac{y}{y^2-y_y^2},
\end{equation}
which integrates to~\eqref{eq:firstintegral}, since $y_\theta^2/y_y^2=6\sigma/(1+4\sigma)$. For the slip wall, $b(y)/a(y)=\mathcal{F}_\theta(\tilde{y};\tilde{\epsilon})/[\lambda\mathcal{F}_y(\tilde{y};\tilde{\epsilon})]$, so the conserved quantity is
\begin{equation}
H(\tilde{y},\theta)=\frac{\sin2\theta}{\Phi(\tilde{y})},\qquad \Phi(\tilde{y})=\exp\left[2\int_{\tilde{y}_\theta}^{\tilde{y}}\frac{\mathcal{F}_\theta(s;\tilde\epsilon)}{\mathcal{F}_y(s;\tilde\epsilon)}\,\dd s\right].
\label{eq:firstintegral_slip}
\end{equation}
For $\lambda<\lambda_c$, $\Phi(\tilde{y})$ has its minimum $\Phi(\tilde{y})=1$ at $\tilde{y}_\theta$ and grows without bound as $\tilde{y}\to\infty$ and as $\tilde{y}\to\tilde{y}_y^+$, where $\Phi(\tilde{y})\sim(\tilde{y}-\tilde{y}_y)^{-2c}$ with $c=-\mathcal{F}_\theta(\tilde{y}_y;\tilde{\epsilon})/\mathcal{F}_y'(\tilde{y}_y;\tilde{\epsilon})>0$; each level set $0<H(\tilde{y},\theta)<1$ is therefore a closed orbit in $\tilde{y}>\tilde{y}_y$.


\begin{thebibliography}{62}%
\makeatletter
\providecommand \@ifxundefined [1]{%
 \@ifx{#1\undefined}
}%
\providecommand \@ifnum [1]{%
 \ifnum #1\expandafter \@firstoftwo
 \else \expandafter \@secondoftwo
 \fi
}%
\providecommand \@ifx [1]{%
 \ifx #1\expandafter \@firstoftwo
 \else \expandafter \@secondoftwo
 \fi
}%
\providecommand \natexlab [1]{#1}%
\providecommand \enquote  [1]{``#1''}%
\providecommand \bibnamefont  [1]{#1}%
\providecommand \bibfnamefont [1]{#1}%
\providecommand \citenamefont [1]{#1}%
\providecommand \href@noop [0]{\@secondoftwo}%
\providecommand \href [0]{\begingroup \@sanitize@url \@href}%
\providecommand \@href[1]{\@@startlink{#1}\@@href}%
\providecommand \@@href[1]{\endgroup#1\@@endlink}%
\providecommand \@sanitize@url [0]{\catcode `\\12\catcode `\$12\catcode `\&12\catcode `\#12\catcode `\^12\catcode `\_12\catcode `\%12\relax}%
\providecommand \@@startlink[1]{}%
\providecommand \@@endlink[0]{}%
\providecommand \url  [0]{\begingroup\@sanitize@url \@url }%
\providecommand \@url [1]{\endgroup\@href {#1}{\urlprefix }}%
\providecommand \urlprefix  [0]{URL }%
\providecommand \Eprint [0]{\href }%
\providecommand \doibase [0]{https://doi.org/}%
\providecommand \selectlanguage [0]{\@gobble}%
\providecommand \bibinfo  [0]{\@secondoftwo}%
\providecommand \bibfield  [0]{\@secondoftwo}%
\providecommand \translation [1]{[#1]}%
\providecommand \BibitemOpen [0]{}%
\providecommand \bibitemStop [0]{}%
\providecommand \bibitemNoStop [0]{.\EOS\space}%
\providecommand \EOS [0]{\spacefactor3000\relax}%
\providecommand \BibitemShut  [1]{\csname bibitem#1\endcsname}%
\let\auto@bib@innerbib\@empty
\bibitem [{\citenamefont {Lauga}\ \emph {et~al.}(2006)\citenamefont {Lauga}, \citenamefont {DiLuzio}, \citenamefont {Whitesides},\ and\ \citenamefont {Stone}}]{lauga2006swimming}%
  \BibitemOpen
  \bibfield  {author} {\bibinfo {author} {\bibfnamefont {E.}~\bibnamefont {Lauga}}, \bibinfo {author} {\bibfnamefont {W.~R.}\ \bibnamefont {DiLuzio}}, \bibinfo {author} {\bibfnamefont {G.~M.}\ \bibnamefont {Whitesides}},\ and\ \bibinfo {author} {\bibfnamefont {H.~A.}\ \bibnamefont {Stone}},\ }\bibfield  {title} {\bibinfo {title} {Swimming in circles: motion of bacteria near solid boundaries},\ }\href@noop {} {\bibfield  {journal} {\bibinfo  {journal} {Biophysical Journal}\ }\textbf {\bibinfo {volume} {90}},\ \bibinfo {pages} {400} (\bibinfo {year} {2006})}\BibitemShut {NoStop}%
\bibitem [{\citenamefont {Berke}\ \emph {et~al.}(2008)\citenamefont {Berke}, \citenamefont {Turner}, \citenamefont {Berg},\ and\ \citenamefont {Lauga}}]{berke2008hydrodynamic}%
  \BibitemOpen
  \bibfield  {author} {\bibinfo {author} {\bibfnamefont {A.~P.}\ \bibnamefont {Berke}}, \bibinfo {author} {\bibfnamefont {L.}~\bibnamefont {Turner}}, \bibinfo {author} {\bibfnamefont {H.~C.}\ \bibnamefont {Berg}},\ and\ \bibinfo {author} {\bibfnamefont {E.}~\bibnamefont {Lauga}},\ }\bibfield  {title} {\bibinfo {title} {Hydrodynamic attraction of swimming microorganisms by surfaces},\ }\href@noop {} {\bibfield  {journal} {\bibinfo  {journal} {Physical Review Letters}\ }\textbf {\bibinfo {volume} {101}},\ \bibinfo {pages} {038102} (\bibinfo {year} {2008})}\BibitemShut {NoStop}%
\bibitem [{\citenamefont {Lauga}\ and\ \citenamefont {Powers}(2009)}]{lauga2009hydrodynamics}%
  \BibitemOpen
  \bibfield  {author} {\bibinfo {author} {\bibfnamefont {E.}~\bibnamefont {Lauga}}\ and\ \bibinfo {author} {\bibfnamefont {T.~R.}\ \bibnamefont {Powers}},\ }\bibfield  {title} {\bibinfo {title} {The hydrodynamics of swimming microorganisms},\ }\href@noop {} {\bibfield  {journal} {\bibinfo  {journal} {Reports on Progress in Physics}\ }\textbf {\bibinfo {volume} {72}},\ \bibinfo {pages} {096601} (\bibinfo {year} {2009})}\BibitemShut {NoStop}%
\bibitem [{\citenamefont {Di~Leonardo}\ \emph {et~al.}(2011)\citenamefont {Di~Leonardo}, \citenamefont {Dell’Arciprete}, \citenamefont {Angelani},\ and\ \citenamefont {Iebba}}]{di2011swimming}%
  \BibitemOpen
  \bibfield  {author} {\bibinfo {author} {\bibfnamefont {R.}~\bibnamefont {Di~Leonardo}}, \bibinfo {author} {\bibfnamefont {D.}~\bibnamefont {Dell’Arciprete}}, \bibinfo {author} {\bibfnamefont {L.}~\bibnamefont {Angelani}},\ and\ \bibinfo {author} {\bibfnamefont {V.}~\bibnamefont {Iebba}},\ }\bibfield  {title} {\bibinfo {title} {Swimming with an image},\ }\href@noop {} {\bibfield  {journal} {\bibinfo  {journal} {Physical Review Letters}\ }\textbf {\bibinfo {volume} {106}},\ \bibinfo {pages} {038101} (\bibinfo {year} {2011})}\BibitemShut {NoStop}%
\bibitem [{\citenamefont {Spagnolie}\ and\ \citenamefont {Lauga}(2012)}]{spagnolie2012hydrodynamics}%
  \BibitemOpen
  \bibfield  {author} {\bibinfo {author} {\bibfnamefont {S.~E.}\ \bibnamefont {Spagnolie}}\ and\ \bibinfo {author} {\bibfnamefont {E.}~\bibnamefont {Lauga}},\ }\bibfield  {title} {\bibinfo {title} {Hydrodynamics of self-propulsion near a boundary: predictions and accuracy of far-field approximations},\ }\href@noop {} {\bibfield  {journal} {\bibinfo  {journal} {Journal of Fluid Mechanics}\ }\textbf {\bibinfo {volume} {700}},\ \bibinfo {pages} {105} (\bibinfo {year} {2012})}\BibitemShut {NoStop}%
\bibitem [{\citenamefont {Elgeti}\ and\ \citenamefont {Gompper}(2016)}]{elgeti2016microswimmers}%
  \BibitemOpen
  \bibfield  {author} {\bibinfo {author} {\bibfnamefont {J.}~\bibnamefont {Elgeti}}\ and\ \bibinfo {author} {\bibfnamefont {G.}~\bibnamefont {Gompper}},\ }\bibfield  {title} {\bibinfo {title} {Microswimmers near surfaces},\ }\href@noop {} {\bibfield  {journal} {\bibinfo  {journal} {The European Physical Journal Special Topics}\ }\textbf {\bibinfo {volume} {225}},\ \bibinfo {pages} {2333} (\bibinfo {year} {2016})}\BibitemShut {NoStop}%
\bibitem [{\citenamefont {Bechinger}\ \emph {et~al.}(2016)\citenamefont {Bechinger}, \citenamefont {Di~Leonardo}, \citenamefont {L{\"o}wen}, \citenamefont {Reichhardt}, \citenamefont {Volpe},\ and\ \citenamefont {Volpe}}]{bechinger2016active}%
  \BibitemOpen
  \bibfield  {author} {\bibinfo {author} {\bibfnamefont {C.}~\bibnamefont {Bechinger}}, \bibinfo {author} {\bibfnamefont {R.}~\bibnamefont {Di~Leonardo}}, \bibinfo {author} {\bibfnamefont {H.}~\bibnamefont {L{\"o}wen}}, \bibinfo {author} {\bibfnamefont {C.}~\bibnamefont {Reichhardt}}, \bibinfo {author} {\bibfnamefont {G.}~\bibnamefont {Volpe}},\ and\ \bibinfo {author} {\bibfnamefont {G.}~\bibnamefont {Volpe}},\ }\bibfield  {title} {\bibinfo {title} {Active particles in complex and crowded environments},\ }\href@noop {} {\bibfield  {journal} {\bibinfo  {journal} {Reviews of Modern Physics}\ }\textbf {\bibinfo {volume} {88}},\ \bibinfo {pages} {045006} (\bibinfo {year} {2016})}\BibitemShut {NoStop}%
\bibitem [{\citenamefont {Ishimoto}\ and\ \citenamefont {Gaffney}(2013)}]{ishimoto2013squirmer}%
  \BibitemOpen
  \bibfield  {author} {\bibinfo {author} {\bibfnamefont {K.}~\bibnamefont {Ishimoto}}\ and\ \bibinfo {author} {\bibfnamefont {E.~A.}\ \bibnamefont {Gaffney}},\ }\bibfield  {title} {\bibinfo {title} {Squirmer dynamics near a boundary},\ }\href@noop {} {\bibfield  {journal} {\bibinfo  {journal} {Physical Review E}\ }\textbf {\bibinfo {volume} {88}},\ \bibinfo {pages} {062702} (\bibinfo {year} {2013})}\BibitemShut {NoStop}%
\bibitem [{\citenamefont {Li}\ and\ \citenamefont {Ardekani}(2014)}]{li2014hydrodynamic}%
  \BibitemOpen
  \bibfield  {author} {\bibinfo {author} {\bibfnamefont {G.-J.}\ \bibnamefont {Li}}\ and\ \bibinfo {author} {\bibfnamefont {A.~M.}\ \bibnamefont {Ardekani}},\ }\bibfield  {title} {\bibinfo {title} {Hydrodynamic interaction of microswimmers near a wall},\ }\href@noop {} {\bibfield  {journal} {\bibinfo  {journal} {Physical Review E}\ }\textbf {\bibinfo {volume} {90}},\ \bibinfo {pages} {013010} (\bibinfo {year} {2014})}\BibitemShut {NoStop}%
\bibitem [{\citenamefont {Uspal}\ \emph {et~al.}(2015)\citenamefont {Uspal}, \citenamefont {Popescu}, \citenamefont {Dietrich},\ and\ \citenamefont {Tasinkevych}}]{uspal2015self}%
  \BibitemOpen
  \bibfield  {author} {\bibinfo {author} {\bibfnamefont {W.~E.}\ \bibnamefont {Uspal}}, \bibinfo {author} {\bibfnamefont {M.~N.}\ \bibnamefont {Popescu}}, \bibinfo {author} {\bibfnamefont {S.}~\bibnamefont {Dietrich}},\ and\ \bibinfo {author} {\bibfnamefont {M.}~\bibnamefont {Tasinkevych}},\ }\bibfield  {title} {\bibinfo {title} {Self-propulsion of a catalytically active particle near a planar wall: from reflection to sliding and hovering},\ }\href@noop {} {\bibfield  {journal} {\bibinfo  {journal} {Soft Matter}\ }\textbf {\bibinfo {volume} {11}},\ \bibinfo {pages} {434} (\bibinfo {year} {2015})}\BibitemShut {NoStop}%
\bibitem [{\citenamefont {Htet}\ \emph {et~al.}(2024)\citenamefont {Htet}, \citenamefont {Das},\ and\ \citenamefont {Lauga}}]{htet2024hydrodynamic}%
  \BibitemOpen
  \bibfield  {author} {\bibinfo {author} {\bibfnamefont {P.~H.}\ \bibnamefont {Htet}}, \bibinfo {author} {\bibfnamefont {D.}~\bibnamefont {Das}},\ and\ \bibinfo {author} {\bibfnamefont {E.}~\bibnamefont {Lauga}},\ }\bibfield  {title} {\bibinfo {title} {Hydrodynamic hovering of swimming bacteria above surfaces},\ }\href@noop {} {\bibfield  {journal} {\bibinfo  {journal} {Physical Review Research}\ }\textbf {\bibinfo {volume} {6}},\ \bibinfo {pages} {L032070} (\bibinfo {year} {2024})}\BibitemShut {NoStop}%
\bibitem [{\citenamefont {Lauga}\ and\ \citenamefont {Squires}(2005)}]{lauga2005brownian}%
  \BibitemOpen
  \bibfield  {author} {\bibinfo {author} {\bibfnamefont {E.}~\bibnamefont {Lauga}}\ and\ \bibinfo {author} {\bibfnamefont {T.~M.}\ \bibnamefont {Squires}},\ }\bibfield  {title} {\bibinfo {title} {Brownian motion near a partial-slip boundary: A local probe of the no-slip condition},\ }\href@noop {} {\bibfield  {journal} {\bibinfo  {journal} {Physics of Fluids}\ }\textbf {\bibinfo {volume} {17}},\ \bibinfo {pages} {103102} (\bibinfo {year} {2005})}\BibitemShut {NoStop}%
\bibitem [{\citenamefont {Lemelle}\ \emph {et~al.}(2013)\citenamefont {Lemelle}, \citenamefont {Palierne}, \citenamefont {Chatre}, \citenamefont {Vaillant},\ and\ \citenamefont {Place}}]{lemelle2013curvature}%
  \BibitemOpen
  \bibfield  {author} {\bibinfo {author} {\bibfnamefont {L.}~\bibnamefont {Lemelle}}, \bibinfo {author} {\bibfnamefont {J.-F.}\ \bibnamefont {Palierne}}, \bibinfo {author} {\bibfnamefont {E.}~\bibnamefont {Chatre}}, \bibinfo {author} {\bibfnamefont {C.}~\bibnamefont {Vaillant}},\ and\ \bibinfo {author} {\bibfnamefont {C.}~\bibnamefont {Place}},\ }\bibfield  {title} {\bibinfo {title} {Curvature reversal of the circular motion of swimming bacteria probes for slip at solid/liquid interfaces},\ }\href@noop {} {\bibfield  {journal} {\bibinfo  {journal} {Soft Matter}\ }\textbf {\bibinfo {volume} {9}},\ \bibinfo {pages} {9759} (\bibinfo {year} {2013})}\BibitemShut {NoStop}%
\bibitem [{\citenamefont {Hu}\ \emph {et~al.}(2015)\citenamefont {Hu}, \citenamefont {Wysocki}, \citenamefont {Winkler},\ and\ \citenamefont {Gompper}}]{hu2015physical}%
  \BibitemOpen
  \bibfield  {author} {\bibinfo {author} {\bibfnamefont {J.}~\bibnamefont {Hu}}, \bibinfo {author} {\bibfnamefont {A.}~\bibnamefont {Wysocki}}, \bibinfo {author} {\bibfnamefont {R.~G.}\ \bibnamefont {Winkler}},\ and\ \bibinfo {author} {\bibfnamefont {G.}~\bibnamefont {Gompper}},\ }\bibfield  {title} {\bibinfo {title} {Physical sensing of surface properties by microswimmers--directing bacterial motion via wall slip},\ }\href@noop {} {\bibfield  {journal} {\bibinfo  {journal} {Scientific Reports}\ }\textbf {\bibinfo {volume} {5}},\ \bibinfo {pages} {9586} (\bibinfo {year} {2015})}\BibitemShut {NoStop}%
\bibitem [{\citenamefont {Ketzetzi}\ \emph {et~al.}(2020)\citenamefont {Ketzetzi}, \citenamefont {De~Graaf}, \citenamefont {Doherty},\ and\ \citenamefont {Kraft}}]{ketzetzi2020slip}%
  \BibitemOpen
  \bibfield  {author} {\bibinfo {author} {\bibfnamefont {S.}~\bibnamefont {Ketzetzi}}, \bibinfo {author} {\bibfnamefont {J.}~\bibnamefont {De~Graaf}}, \bibinfo {author} {\bibfnamefont {R.~P.}\ \bibnamefont {Doherty}},\ and\ \bibinfo {author} {\bibfnamefont {D.~J.}\ \bibnamefont {Kraft}},\ }\bibfield  {title} {\bibinfo {title} {Slip length dependent propulsion speed of catalytic colloidal swimmers near walls},\ }\href@noop {} {\bibfield  {journal} {\bibinfo  {journal} {Physical Review Letters}\ }\textbf {\bibinfo {volume} {124}},\ \bibinfo {pages} {048002} (\bibinfo {year} {2020})}\BibitemShut {NoStop}%
\bibitem [{\citenamefont {Fulford}\ and\ \citenamefont {Blake}(1983)}]{fulford1983motion}%
  \BibitemOpen
  \bibfield  {author} {\bibinfo {author} {\bibfnamefont {G.~R.}\ \bibnamefont {Fulford}}\ and\ \bibinfo {author} {\bibfnamefont {J.~R.}\ \bibnamefont {Blake}},\ }\bibfield  {title} {\bibinfo {title} {On the motion of a slender body near an interface between two immiscible liquids at very low {R}eynolds numbers},\ }\href@noop {} {\bibfield  {journal} {\bibinfo  {journal} {Journal of Fluid Mechanics}\ }\textbf {\bibinfo {volume} {127}},\ \bibinfo {pages} {203} (\bibinfo {year} {1983})}\BibitemShut {NoStop}%
\bibitem [{\citenamefont {Lopez}\ and\ \citenamefont {Lauga}(2014)}]{lopez2014dynamics}%
  \BibitemOpen
  \bibfield  {author} {\bibinfo {author} {\bibfnamefont {D.}~\bibnamefont {Lopez}}\ and\ \bibinfo {author} {\bibfnamefont {E.}~\bibnamefont {Lauga}},\ }\bibfield  {title} {\bibinfo {title} {Dynamics of swimming bacteria at complex interfaces},\ }\href@noop {} {\bibfield  {journal} {\bibinfo  {journal} {Physics of Fluids}\ }\textbf {\bibinfo {volume} {26}},\ \bibinfo {pages} {071902} (\bibinfo {year} {2014})}\BibitemShut {NoStop}%
\bibitem [{\citenamefont {Feng}\ \emph {et~al.}(2022)\citenamefont {Feng}, \citenamefont {Molina}, \citenamefont {Turner},\ and\ \citenamefont {Yamamoto}}]{feng2022dynamics}%
  \BibitemOpen
  \bibfield  {author} {\bibinfo {author} {\bibfnamefont {C.}~\bibnamefont {Feng}}, \bibinfo {author} {\bibfnamefont {J.~J.}\ \bibnamefont {Molina}}, \bibinfo {author} {\bibfnamefont {M.~S.}\ \bibnamefont {Turner}},\ and\ \bibinfo {author} {\bibfnamefont {R.}~\bibnamefont {Yamamoto}},\ }\bibfield  {title} {\bibinfo {title} {Dynamics of microswimmers near a soft penetrable interface},\ }\href@noop {} {\bibfield  {journal} {\bibinfo  {journal} {Physical Review Research}\ }\textbf {\bibinfo {volume} {4}},\ \bibinfo {pages} {043202} (\bibinfo {year} {2022})}\BibitemShut {NoStop}%
\bibitem [{\citenamefont {Feng}\ \emph {et~al.}(2023)\citenamefont {Feng}, \citenamefont {Molina}, \citenamefont {Turner},\ and\ \citenamefont {Yamamoto}}]{feng2023dynamics}%
  \BibitemOpen
  \bibfield  {author} {\bibinfo {author} {\bibfnamefont {C.}~\bibnamefont {Feng}}, \bibinfo {author} {\bibfnamefont {J.~J.}\ \bibnamefont {Molina}}, \bibinfo {author} {\bibfnamefont {M.~S.}\ \bibnamefont {Turner}},\ and\ \bibinfo {author} {\bibfnamefont {R.}~\bibnamefont {Yamamoto}},\ }\bibfield  {title} {\bibinfo {title} {Dynamics of microswimmers near a liquid--liquid interface with viscosity difference},\ }\href@noop {} {\bibfield  {journal} {\bibinfo  {journal} {Physics of Fluids}\ }\textbf {\bibinfo {volume} {35}},\ \bibinfo {pages} {051903} (\bibinfo {year} {2023})}\BibitemShut {NoStop}%
\bibitem [{\citenamefont {Poddar}\ \emph {et~al.}(2020)\citenamefont {Poddar}, \citenamefont {Bandopadhyay},\ and\ \citenamefont {Chakraborty}}]{poddar2020near}%
  \BibitemOpen
  \bibfield  {author} {\bibinfo {author} {\bibfnamefont {A.}~\bibnamefont {Poddar}}, \bibinfo {author} {\bibfnamefont {A.}~\bibnamefont {Bandopadhyay}},\ and\ \bibinfo {author} {\bibfnamefont {S.}~\bibnamefont {Chakraborty}},\ }\bibfield  {title} {\bibinfo {title} {Near-wall hydrodynamic slip triggers swimming state transition of micro-organisms},\ }\href@noop {} {\bibfield  {journal} {\bibinfo  {journal} {Journal of Fluid Mechanics}\ }\textbf {\bibinfo {volume} {894}},\ \bibinfo {pages} {A11} (\bibinfo {year} {2020})}\BibitemShut {NoStop}%
\bibitem [{\citenamefont {Ghosh}\ \emph {et~al.}(2026)\citenamefont {Ghosh}, \citenamefont {Mandal},\ and\ \citenamefont {Poddar}}]{ghosh2026slippery}%
  \BibitemOpen
  \bibfield  {author} {\bibinfo {author} {\bibfnamefont {S.}~\bibnamefont {Ghosh}}, \bibinfo {author} {\bibfnamefont {S.}~\bibnamefont {Mandal}},\ and\ \bibinfo {author} {\bibfnamefont {A.}~\bibnamefont {Poddar}},\ }\bibfield  {title} {\bibinfo {title} {Slippery boundaries enhance trapping of catalytic micromotors},\ }\href@noop {} {\bibfield  {journal} {\bibinfo  {journal} {Journal of Fluid Mechanics}\ }\textbf {\bibinfo {volume} {1044}},\ \bibinfo {pages} {A5} (\bibinfo {year} {2026})}\BibitemShut {NoStop}%
\bibitem [{\citenamefont {Ghosh}\ and\ \citenamefont {Poddar}(2023)}]{ghosh2023slippery}%
  \BibitemOpen
  \bibfield  {author} {\bibinfo {author} {\bibfnamefont {S.}~\bibnamefont {Ghosh}}\ and\ \bibinfo {author} {\bibfnamefont {A.}~\bibnamefont {Poddar}},\ }\bibfield  {title} {\bibinfo {title} {Slippery rheotaxis: new regimes for guiding wall-bound microswimmers},\ }\href@noop {} {\bibfield  {journal} {\bibinfo  {journal} {Journal of Fluid Mechanics}\ }\textbf {\bibinfo {volume} {967}},\ \bibinfo {pages} {A14} (\bibinfo {year} {2023})}\BibitemShut {NoStop}%
\bibitem [{\citenamefont {Ghosh}\ \emph {et~al.}(2024)\citenamefont {Ghosh}, \citenamefont {Ghoshal},\ and\ \citenamefont {Poddar}}]{ghosh2024harnessing}%
  \BibitemOpen
  \bibfield  {author} {\bibinfo {author} {\bibfnamefont {S.}~\bibnamefont {Ghosh}}, \bibinfo {author} {\bibfnamefont {S.}~\bibnamefont {Ghoshal}},\ and\ \bibinfo {author} {\bibfnamefont {A.}~\bibnamefont {Poddar}},\ }\bibfield  {title} {\bibinfo {title} {Harnessing wall slip towards tunable microswimming in poiseuille flow},\ }\href@noop {} {\bibfield  {journal} {\bibinfo  {journal} {Journal of Fluid Mechanics}\ }\textbf {\bibinfo {volume} {997}},\ \bibinfo {pages} {A59} (\bibinfo {year} {2024})}\BibitemShut {NoStop}%
\bibitem [{\citenamefont {Guan}\ \emph {et~al.}(2024)\citenamefont {Guan}, \citenamefont {Ying},\ and\ \citenamefont {Lin}}]{guan2024effect}%
  \BibitemOpen
  \bibfield  {author} {\bibinfo {author} {\bibfnamefont {G.}~\bibnamefont {Guan}}, \bibinfo {author} {\bibfnamefont {Y.}~\bibnamefont {Ying}},\ and\ \bibinfo {author} {\bibfnamefont {J.}~\bibnamefont {Lin}},\ }\bibfield  {title} {\bibinfo {title} {Effect of the shape of microswimmers and slip boundary conditions on the dynamic characteristics of near-wall microswimmers},\ }\href@noop {} {\bibfield  {journal} {\bibinfo  {journal} {Journal of Fluid Mechanics}\ }\textbf {\bibinfo {volume} {999}},\ \bibinfo {pages} {A42} (\bibinfo {year} {2024})}\BibitemShut {NoStop}%
\bibitem [{\citenamefont {Kamal}\ \emph {et~al.}(2020)\citenamefont {Kamal}, \citenamefont {Gravelle},\ and\ \citenamefont {Botto}}]{kamal2020hydrodynamic}%
  \BibitemOpen
  \bibfield  {author} {\bibinfo {author} {\bibfnamefont {C.}~\bibnamefont {Kamal}}, \bibinfo {author} {\bibfnamefont {S.}~\bibnamefont {Gravelle}},\ and\ \bibinfo {author} {\bibfnamefont {L.}~\bibnamefont {Botto}},\ }\bibfield  {title} {\bibinfo {title} {Hydrodynamic slip can align thin nanoplatelets in shear flow},\ }\href@noop {} {\bibfield  {journal} {\bibinfo  {journal} {Nature Communications}\ }\textbf {\bibinfo {volume} {11}},\ \bibinfo {pages} {2425} (\bibinfo {year} {2020})}\BibitemShut {NoStop}%
\bibitem [{\citenamefont {Gravelle}\ \emph {et~al.}(2021)\citenamefont {Gravelle}, \citenamefont {Kamal},\ and\ \citenamefont {Botto}}]{gravelle2021violations}%
  \BibitemOpen
  \bibfield  {author} {\bibinfo {author} {\bibfnamefont {S.}~\bibnamefont {Gravelle}}, \bibinfo {author} {\bibfnamefont {C.}~\bibnamefont {Kamal}},\ and\ \bibinfo {author} {\bibfnamefont {L.}~\bibnamefont {Botto}},\ }\bibfield  {title} {\bibinfo {title} {Violations of {J}effery's theory in the dynamics of nanographene in shear flow},\ }\href@noop {} {\bibfield  {journal} {\bibinfo  {journal} {Physical Review Fluids}\ }\textbf {\bibinfo {volume} {6}},\ \bibinfo {pages} {034303} (\bibinfo {year} {2021})}\BibitemShut {NoStop}%
\bibitem [{\citenamefont {Crowdy}(2022)}]{crowdy2022equilibrium}%
  \BibitemOpen
  \bibfield  {author} {\bibinfo {author} {\bibfnamefont {D.~G.}\ \bibnamefont {Crowdy}},\ }\bibfield  {title} {\bibinfo {title} {Equilibrium tilt of slippery elliptical rods in creeping simple shear},\ }\href@noop {} {\bibfield  {journal} {\bibinfo  {journal} {Journal of Fluid Mechanics}\ }\textbf {\bibinfo {volume} {931}},\ \bibinfo {pages} {R2} (\bibinfo {year} {2022})}\BibitemShut {NoStop}%
\bibitem [{\citenamefont {Crowdy}\ and\ \citenamefont {Or}(2010)}]{crowdy2010two}%
  \BibitemOpen
  \bibfield  {author} {\bibinfo {author} {\bibfnamefont {D.~G.}\ \bibnamefont {Crowdy}}\ and\ \bibinfo {author} {\bibfnamefont {Y.}~\bibnamefont {Or}},\ }\bibfield  {title} {\bibinfo {title} {Two-dimensional point singularity model of a low-{R}eynolds-number swimmer near a wall},\ }\href@noop {} {\bibfield  {journal} {\bibinfo  {journal} {Physical Review E}\ }\textbf {\bibinfo {volume} {81}},\ \bibinfo {pages} {036313} (\bibinfo {year} {2010})}\BibitemShut {NoStop}%
\bibitem [{\citenamefont {Crowdy}(2011)}]{crowdy2011treadmilling}%
  \BibitemOpen
  \bibfield  {author} {\bibinfo {author} {\bibfnamefont {D.}~\bibnamefont {Crowdy}},\ }\bibfield  {title} {\bibinfo {title} {Treadmilling swimmers near a no-slip wall at low {R}eynolds number},\ }\href@noop {} {\bibfield  {journal} {\bibinfo  {journal} {International Journal of Non-Linear Mechanics}\ }\textbf {\bibinfo {volume} {46}},\ \bibinfo {pages} {577} (\bibinfo {year} {2011})}\BibitemShut {NoStop}%
\bibitem [{\citenamefont {Ishimoto}\ and\ \citenamefont {Crowdy}(2017)}]{ishimoto2017dynamics}%
  \BibitemOpen
  \bibfield  {author} {\bibinfo {author} {\bibfnamefont {K.}~\bibnamefont {Ishimoto}}\ and\ \bibinfo {author} {\bibfnamefont {D.~G.}\ \bibnamefont {Crowdy}},\ }\bibfield  {title} {\bibinfo {title} {Dynamics of a treadmilling microswimmer near a no-slip wall in simple shear},\ }\href@noop {} {\bibfield  {journal} {\bibinfo  {journal} {Journal of Fluid Mechanics}\ }\textbf {\bibinfo {volume} {821}},\ \bibinfo {pages} {647} (\bibinfo {year} {2017})}\BibitemShut {NoStop}%
\bibitem [{\citenamefont {Crowdy}\ \emph {et~al.}(2011)\citenamefont {Crowdy}, \citenamefont {Lee}, \citenamefont {Samson}, \citenamefont {Lauga},\ and\ \citenamefont {Hosoi}}]{crowdy2011two}%
  \BibitemOpen
  \bibfield  {author} {\bibinfo {author} {\bibfnamefont {D.}~\bibnamefont {Crowdy}}, \bibinfo {author} {\bibfnamefont {S.}~\bibnamefont {Lee}}, \bibinfo {author} {\bibfnamefont {O.}~\bibnamefont {Samson}}, \bibinfo {author} {\bibfnamefont {E.}~\bibnamefont {Lauga}},\ and\ \bibinfo {author} {\bibfnamefont {A.}~\bibnamefont {Hosoi}},\ }\bibfield  {title} {\bibinfo {title} {A two-dimensional model of low-{R}eynolds number swimming beneath a free surface},\ }\href@noop {} {\bibfield  {journal} {\bibinfo  {journal} {Journal of Fluid Mechanics}\ }\textbf {\bibinfo {volume} {681}},\ \bibinfo {pages} {24} (\bibinfo {year} {2011})}\BibitemShut {NoStop}%
\bibitem [{\citenamefont {Crowdy}\ and\ \citenamefont {Samson}(2011)}]{crowdy2011hydrodynamic}%
  \BibitemOpen
  \bibfield  {author} {\bibinfo {author} {\bibfnamefont {D.}~\bibnamefont {Crowdy}}\ and\ \bibinfo {author} {\bibfnamefont {O.}~\bibnamefont {Samson}},\ }\bibfield  {title} {\bibinfo {title} {Hydrodynamic bound states of a low-{R}eynolds-number swimmer near a gap in a wall},\ }\href@noop {} {\bibfield  {journal} {\bibinfo  {journal} {Journal of Fluid Mechanics}\ }\textbf {\bibinfo {volume} {667}},\ \bibinfo {pages} {309} (\bibinfo {year} {2011})}\BibitemShut {NoStop}%
\bibitem [{\citenamefont {Davis}\ and\ \citenamefont {Crowdy}(2012)}]{davis2012stresslet}%
  \BibitemOpen
  \bibfield  {author} {\bibinfo {author} {\bibfnamefont {A.~M.~J.}\ \bibnamefont {Davis}}\ and\ \bibinfo {author} {\bibfnamefont {D.~G.}\ \bibnamefont {Crowdy}},\ }\bibfield  {title} {\bibinfo {title} {Stresslet asymptotics for a treadmilling swimmer near a two-dimensional corner: hydrodynamic bound states},\ }\href@noop {} {\bibfield  {journal} {\bibinfo  {journal} {Proceedings of the Royal Society A}\ }\textbf {\bibinfo {volume} {468}},\ \bibinfo {pages} {3765} (\bibinfo {year} {2012})}\BibitemShut {NoStop}%
\bibitem [{\citenamefont {Davis}\ and\ \citenamefont {Crowdy}(2013)}]{davis2013matched}%
  \BibitemOpen
  \bibfield  {author} {\bibinfo {author} {\bibfnamefont {A.~M.~J.}\ \bibnamefont {Davis}}\ and\ \bibinfo {author} {\bibfnamefont {D.~G.}\ \bibnamefont {Crowdy}},\ }\bibfield  {title} {\bibinfo {title} {Matched asymptotics for a treadmilling low-{R}eynolds-number swimmer near a wall},\ }\href@noop {} {\bibfield  {journal} {\bibinfo  {journal} {Quarterly Journal of Mechanics and Applied Mathematics}\ }\textbf {\bibinfo {volume} {66}},\ \bibinfo {pages} {53} (\bibinfo {year} {2013})}\BibitemShut {NoStop}%
\bibitem [{\citenamefont {Or}\ and\ \citenamefont {Murray}(2009)}]{or2009dynamics}%
  \BibitemOpen
  \bibfield  {author} {\bibinfo {author} {\bibfnamefont {Y.}~\bibnamefont {Or}}\ and\ \bibinfo {author} {\bibfnamefont {R.~M.}\ \bibnamefont {Murray}},\ }\bibfield  {title} {\bibinfo {title} {Dynamics and stability of a class of low reynolds number swimmers near a wall},\ }\href@noop {} {\bibfield  {journal} {\bibinfo  {journal} {Physical Review E}\ }\textbf {\bibinfo {volume} {79}},\ \bibinfo {pages} {045302} (\bibinfo {year} {2009})}\BibitemShut {NoStop}%
\bibitem [{\citenamefont {Or}\ \emph {et~al.}(2011)\citenamefont {Or}, \citenamefont {Zhang},\ and\ \citenamefont {Murray}}]{or2011dynamics}%
  \BibitemOpen
  \bibfield  {author} {\bibinfo {author} {\bibfnamefont {Y.}~\bibnamefont {Or}}, \bibinfo {author} {\bibfnamefont {S.}~\bibnamefont {Zhang}},\ and\ \bibinfo {author} {\bibfnamefont {R.~M.}\ \bibnamefont {Murray}},\ }\bibfield  {title} {\bibinfo {title} {Dynamics and stability of low-{R}eynolds-number swimming near a wall},\ }\href@noop {} {\bibfield  {journal} {\bibinfo  {journal} {SIAM Journal on Applied Dynamical Systems}\ }\textbf {\bibinfo {volume} {10}},\ \bibinfo {pages} {1013} (\bibinfo {year} {2011})}\BibitemShut {NoStop}%
\bibitem [{\citenamefont {Zhang}\ \emph {et~al.}(2010)\citenamefont {Zhang}, \citenamefont {Or},\ and\ \citenamefont {Murray}}]{zhang2010experimental}%
  \BibitemOpen
  \bibfield  {author} {\bibinfo {author} {\bibfnamefont {S.}~\bibnamefont {Zhang}}, \bibinfo {author} {\bibfnamefont {Y.}~\bibnamefont {Or}},\ and\ \bibinfo {author} {\bibfnamefont {R.~M.}\ \bibnamefont {Murray}},\ }\bibfield  {title} {\bibinfo {title} {Experimental demonstration of the dynamics and stability of a low {R}eynolds number swimmer near a plane wall},\ }in\ \href@noop {} {\emph {\bibinfo {booktitle} {Proceedings of the 2010 American Control Conference}}}\ (\bibinfo {organization} {IEEE},\ \bibinfo {year} {2010})\ pp.\ \bibinfo {pages} {4205--4210}\BibitemShut {NoStop}%
\bibitem [{\citenamefont {Ishimoto}(2017)}]{ishimoto2017guidance}%
  \BibitemOpen
  \bibfield  {author} {\bibinfo {author} {\bibfnamefont {K.}~\bibnamefont {Ishimoto}},\ }\bibfield  {title} {\bibinfo {title} {Guidance of microswimmers by wall and flow: thigmotaxis and rheotaxis of unsteady squirmers in two and three dimensions},\ }\href@noop {} {\bibfield  {journal} {\bibinfo  {journal} {Physical Review E}\ }\textbf {\bibinfo {volume} {96}},\ \bibinfo {pages} {043103} (\bibinfo {year} {2017})}\BibitemShut {NoStop}%
\bibitem [{\citenamefont {Hatwalne}\ \emph {et~al.}(2004)\citenamefont {Hatwalne}, \citenamefont {Ramaswamy}, \citenamefont {Rao},\ and\ \citenamefont {Simha}}]{hatwalne2004rheology}%
  \BibitemOpen
  \bibfield  {author} {\bibinfo {author} {\bibfnamefont {Y.}~\bibnamefont {Hatwalne}}, \bibinfo {author} {\bibfnamefont {S.}~\bibnamefont {Ramaswamy}}, \bibinfo {author} {\bibfnamefont {M.}~\bibnamefont {Rao}},\ and\ \bibinfo {author} {\bibfnamefont {R.~A.}\ \bibnamefont {Simha}},\ }\bibfield  {title} {\bibinfo {title} {Rheology of active-particle suspensions},\ }\href@noop {} {\bibfield  {journal} {\bibinfo  {journal} {Physical Review Letters}\ }\textbf {\bibinfo {volume} {92}},\ \bibinfo {pages} {118101} (\bibinfo {year} {2004})}\BibitemShut {NoStop}%
\bibitem [{\citenamefont {Yazdi}\ \emph {et~al.}(2015)\citenamefont {Yazdi}, \citenamefont {Ardekani},\ and\ \citenamefont {Borhan}}]{yazdi2015swimming}%
  \BibitemOpen
  \bibfield  {author} {\bibinfo {author} {\bibfnamefont {S.}~\bibnamefont {Yazdi}}, \bibinfo {author} {\bibfnamefont {A.~M.}\ \bibnamefont {Ardekani}},\ and\ \bibinfo {author} {\bibfnamefont {A.}~\bibnamefont {Borhan}},\ }\bibfield  {title} {\bibinfo {title} {Swimming dynamics near a wall in a weakly elastic fluid},\ }\href@noop {} {\bibfield  {journal} {\bibinfo  {journal} {Journal of Nonlinear Science}\ }\textbf {\bibinfo {volume} {25}},\ \bibinfo {pages} {1153} (\bibinfo {year} {2015})}\BibitemShut {NoStop}%
\bibitem [{\citenamefont {Aderogba}\ and\ \citenamefont {Blake}(1978)}]{aderogba1978action}%
  \BibitemOpen
  \bibfield  {author} {\bibinfo {author} {\bibfnamefont {K.}~\bibnamefont {Aderogba}}\ and\ \bibinfo {author} {\bibfnamefont {J.}~\bibnamefont {Blake}},\ }\bibfield  {title} {\bibinfo {title} {Action of a force near the planar surface between two semi-infinite immiscible liquids at very low {R}eynolds numbers},\ }\href@noop {} {\bibfield  {journal} {\bibinfo  {journal} {Bulletin of the Australian Mathematical Society}\ }\textbf {\bibinfo {volume} {18}},\ \bibinfo {pages} {345} (\bibinfo {year} {1978})}\BibitemShut {NoStop}%
\bibitem [{\citenamefont {Crowdy}(2020{\natexlab{a}})}]{flotilla}%
  \BibitemOpen
  \bibfield  {author} {\bibinfo {author} {\bibfnamefont {D.}~\bibnamefont {Crowdy}},\ }\bibfield  {title} {\bibinfo {title} {Collective viscous propulsion of a two-dimensional flotilla of {M}arangoni boats},\ }\href@noop {} {\bibfield  {journal} {\bibinfo  {journal} {Physical Review Fluids}\ }\textbf {\bibinfo {volume} {5}},\ \bibinfo {pages} {124004} (\bibinfo {year} {2020}{\natexlab{a}})}\BibitemShut {NoStop}%
\bibitem [{\citenamefont {Crowdy}(2020{\natexlab{b}})}]{crowdy2020solving}%
  \BibitemOpen
  \bibfield  {author} {\bibinfo {author} {\bibfnamefont {D.}~\bibnamefont {Crowdy}},\ }\href@noop {} {\emph {\bibinfo {title} {Solving problems in multiply connected domains}}},\ \bibinfo {series} {CBMS-NSF Regional Conference Series in Applied Mathematics}, Vol.~\bibinfo {volume} {97}\ (\bibinfo  {publisher} {SIAM},\ \bibinfo {address} {Philadelphia},\ \bibinfo {year} {2020})\BibitemShut {NoStop}%
\bibitem [{\citenamefont {Petroff}\ \emph {et~al.}(2015)\citenamefont {Petroff}, \citenamefont {Wu},\ and\ \citenamefont {Libchaber}}]{petroff2015fast}%
  \BibitemOpen
  \bibfield  {author} {\bibinfo {author} {\bibfnamefont {A.~P.}\ \bibnamefont {Petroff}}, \bibinfo {author} {\bibfnamefont {X.-L.}\ \bibnamefont {Wu}},\ and\ \bibinfo {author} {\bibfnamefont {A.}~\bibnamefont {Libchaber}},\ }\bibfield  {title} {\bibinfo {title} {Fast-moving bacteria self-organize into active two-dimensional crystals of rotating cells},\ }\href@noop {} {\bibfield  {journal} {\bibinfo  {journal} {Physical Review Letters}\ }\textbf {\bibinfo {volume} {114}},\ \bibinfo {pages} {158102} (\bibinfo {year} {2015})}\BibitemShut {NoStop}%
\bibitem [{\citenamefont {Petroff}\ and\ \citenamefont {Libchaber}(2018)}]{petroff2018nucleation}%
  \BibitemOpen
  \bibfield  {author} {\bibinfo {author} {\bibfnamefont {A.~P.}\ \bibnamefont {Petroff}}\ and\ \bibinfo {author} {\bibfnamefont {A.}~\bibnamefont {Libchaber}},\ }\bibfield  {title} {\bibinfo {title} {Nucleation of rotating crystals by \textit{Thiovulum majus} bacteria},\ }\href@noop {} {\bibfield  {journal} {\bibinfo  {journal} {New Journal of Physics}\ }\textbf {\bibinfo {volume} {20}},\ \bibinfo {pages} {015007} (\bibinfo {year} {2018})}\BibitemShut {NoStop}%
\bibitem [{\citenamefont {Das}\ and\ \citenamefont {Lauga}(2019)}]{das2019transition}%
  \BibitemOpen
  \bibfield  {author} {\bibinfo {author} {\bibfnamefont {D.}~\bibnamefont {Das}}\ and\ \bibinfo {author} {\bibfnamefont {E.}~\bibnamefont {Lauga}},\ }\bibfield  {title} {\bibinfo {title} {Transition to bound states for bacteria swimming near surfaces},\ }\href@noop {} {\bibfield  {journal} {\bibinfo  {journal} {Physical Review E}\ }\textbf {\bibinfo {volume} {100}},\ \bibinfo {pages} {043117} (\bibinfo {year} {2019})}\BibitemShut {NoStop}%
\bibitem [{\citenamefont {Kuron}\ \emph {et~al.}(2019)\citenamefont {Kuron}, \citenamefont {St{\"a}rk}, \citenamefont {Holm},\ and\ \citenamefont {De~Graaf}}]{kuron2019hydrodynamic}%
  \BibitemOpen
  \bibfield  {author} {\bibinfo {author} {\bibfnamefont {M.}~\bibnamefont {Kuron}}, \bibinfo {author} {\bibfnamefont {P.}~\bibnamefont {St{\"a}rk}}, \bibinfo {author} {\bibfnamefont {C.}~\bibnamefont {Holm}},\ and\ \bibinfo {author} {\bibfnamefont {J.}~\bibnamefont {De~Graaf}},\ }\bibfield  {title} {\bibinfo {title} {Hydrodynamic mobility reversal of squirmers near flat and curved surfaces},\ }\href@noop {} {\bibfield  {journal} {\bibinfo  {journal} {Soft Matter}\ }\textbf {\bibinfo {volume} {15}},\ \bibinfo {pages} {5908} (\bibinfo {year} {2019})}\BibitemShut {NoStop}%
\bibitem [{\citenamefont {Lee}\ and\ \citenamefont {Leal}(1980)}]{lee1980motion}%
  \BibitemOpen
  \bibfield  {author} {\bibinfo {author} {\bibfnamefont {S.~H.}\ \bibnamefont {Lee}}\ and\ \bibinfo {author} {\bibfnamefont {L.~G.}\ \bibnamefont {Leal}},\ }\bibfield  {title} {\bibinfo {title} {Motion of a sphere in the presence of a plane interface. {P}art 2. {A}n exact solution in bipolar co-ordinates},\ }\href@noop {} {\bibfield  {journal} {\bibinfo  {journal} {Journal of Fluid Mechanics}\ }\textbf {\bibinfo {volume} {98}},\ \bibinfo {pages} {193} (\bibinfo {year} {1980})}\BibitemShut {NoStop}%
\bibitem [{\citenamefont {Drescher}\ \emph {et~al.}(2009)\citenamefont {Drescher}, \citenamefont {Leptos}, \citenamefont {Tuval}, \citenamefont {Ishikawa}, \citenamefont {Pedley},\ and\ \citenamefont {Goldstein}}]{drescher2009dancing}%
  \BibitemOpen
  \bibfield  {author} {\bibinfo {author} {\bibfnamefont {K.}~\bibnamefont {Drescher}}, \bibinfo {author} {\bibfnamefont {K.~C.}\ \bibnamefont {Leptos}}, \bibinfo {author} {\bibfnamefont {I.}~\bibnamefont {Tuval}}, \bibinfo {author} {\bibfnamefont {T.}~\bibnamefont {Ishikawa}}, \bibinfo {author} {\bibfnamefont {T.~J.}\ \bibnamefont {Pedley}},\ and\ \bibinfo {author} {\bibfnamefont {R.~E.}\ \bibnamefont {Goldstein}},\ }\bibfield  {title} {\bibinfo {title} {Dancing \textit{Volvox}: hydrodynamic bound states of swimming algae},\ }\href@noop {} {\bibfield  {journal} {\bibinfo  {journal} {Physical Review Letters}\ }\textbf {\bibinfo {volume} {102}},\ \bibinfo {pages} {168101} (\bibinfo {year} {2009})}\BibitemShut {NoStop}%
\bibitem [{\citenamefont {Ahmadzadegan}\ \emph {et~al.}(2019)\citenamefont {Ahmadzadegan}, \citenamefont {Wang}, \citenamefont {Vlachos},\ and\ \citenamefont {Ardekani}}]{ahmadzadegan2019hydrodynamic}%
  \BibitemOpen
  \bibfield  {author} {\bibinfo {author} {\bibfnamefont {A.}~\bibnamefont {Ahmadzadegan}}, \bibinfo {author} {\bibfnamefont {S.}~\bibnamefont {Wang}}, \bibinfo {author} {\bibfnamefont {P.~P.}\ \bibnamefont {Vlachos}},\ and\ \bibinfo {author} {\bibfnamefont {A.~M.}\ \bibnamefont {Ardekani}},\ }\bibfield  {title} {\bibinfo {title} {Hydrodynamic attraction of bacteria to gas and liquid interfaces},\ }\href@noop {} {\bibfield  {journal} {\bibinfo  {journal} {Physical Review E}\ }\textbf {\bibinfo {volume} {100}},\ \bibinfo {pages} {062605} (\bibinfo {year} {2019})}\BibitemShut {NoStop}%
\bibitem [{\citenamefont {Mathijssen}\ \emph {et~al.}(2016{\natexlab{a}})\citenamefont {Mathijssen}, \citenamefont {Shendruk}, \citenamefont {Yeomans},\ and\ \citenamefont {Doostmohammadi}}]{mathijssen2016upstream}%
  \BibitemOpen
  \bibfield  {author} {\bibinfo {author} {\bibfnamefont {A.~J. T.~M.}\ \bibnamefont {Mathijssen}}, \bibinfo {author} {\bibfnamefont {T.~N.}\ \bibnamefont {Shendruk}}, \bibinfo {author} {\bibfnamefont {J.~M.}\ \bibnamefont {Yeomans}},\ and\ \bibinfo {author} {\bibfnamefont {A.}~\bibnamefont {Doostmohammadi}},\ }\bibfield  {title} {\bibinfo {title} {Upstream swimming in microbiological flows},\ }\href@noop {} {\bibfield  {journal} {\bibinfo  {journal} {Physical Review Letters}\ }\textbf {\bibinfo {volume} {116}},\ \bibinfo {pages} {028104} (\bibinfo {year} {2016}{\natexlab{a}})}\BibitemShut {NoStop}%
\bibitem [{\citenamefont {Zaferani}\ \emph {et~al.}(2018)\citenamefont {Zaferani}, \citenamefont {Cheong},\ and\ \citenamefont {Abbaspourrad}}]{zaferani2018rheotaxis}%
  \BibitemOpen
  \bibfield  {author} {\bibinfo {author} {\bibfnamefont {M.}~\bibnamefont {Zaferani}}, \bibinfo {author} {\bibfnamefont {S.~H.}\ \bibnamefont {Cheong}},\ and\ \bibinfo {author} {\bibfnamefont {A.}~\bibnamefont {Abbaspourrad}},\ }\bibfield  {title} {\bibinfo {title} {Rheotaxis-based separation of sperm with progressive motility using a microfluidic corral system},\ }\href@noop {} {\bibfield  {journal} {\bibinfo  {journal} {Proceedings of the National Academy of Sciences}\ }\textbf {\bibinfo {volume} {115}},\ \bibinfo {pages} {8272} (\bibinfo {year} {2018})}\BibitemShut {NoStop}%
\bibitem [{\citenamefont {Lemelle}\ \emph {et~al.}(2010)\citenamefont {Lemelle}, \citenamefont {Palierne}, \citenamefont {Chatre},\ and\ \citenamefont {Place}}]{lemelle2010counterclockwise}%
  \BibitemOpen
  \bibfield  {author} {\bibinfo {author} {\bibfnamefont {L.}~\bibnamefont {Lemelle}}, \bibinfo {author} {\bibfnamefont {J.-F.}\ \bibnamefont {Palierne}}, \bibinfo {author} {\bibfnamefont {E.}~\bibnamefont {Chatre}},\ and\ \bibinfo {author} {\bibfnamefont {C.}~\bibnamefont {Place}},\ }\bibfield  {title} {\bibinfo {title} {Counterclockwise circular motion of bacteria swimming at the air-liquid interface},\ }\href@noop {} {\bibfield  {journal} {\bibinfo  {journal} {Journal of Bacteriology}\ }\textbf {\bibinfo {volume} {192}},\ \bibinfo {pages} {6307} (\bibinfo {year} {2010})}\BibitemShut {NoStop}%
\bibitem [{\citenamefont {Guan}\ \emph {et~al.}(2025)\citenamefont {Guan}, \citenamefont {Ying}, \citenamefont {Jiang},\ and\ \citenamefont {Lin}}]{guan2025study}%
  \BibitemOpen
  \bibfield  {author} {\bibinfo {author} {\bibfnamefont {G.}~\bibnamefont {Guan}}, \bibinfo {author} {\bibfnamefont {Y.}~\bibnamefont {Ying}}, \bibinfo {author} {\bibfnamefont {T.}~\bibnamefont {Jiang}},\ and\ \bibinfo {author} {\bibfnamefont {J.}~\bibnamefont {Lin}},\ }\bibfield  {title} {\bibinfo {title} {Study on the motion characteristics of chiral microswimmers near the wall with slip boundary conditions},\ }\href@noop {} {\bibfield  {journal} {\bibinfo  {journal} {Physics of Fluids}\ }\textbf {\bibinfo {volume} {37}},\ \bibinfo {pages} {113331} (\bibinfo {year} {2025})}\BibitemShut {NoStop}%
\bibitem [{\citenamefont {Shaik}\ and\ \citenamefont {Ardekani}(2017)}]{shaik2017motion}%
  \BibitemOpen
  \bibfield  {author} {\bibinfo {author} {\bibfnamefont {V.~A.}\ \bibnamefont {Shaik}}\ and\ \bibinfo {author} {\bibfnamefont {A.~M.}\ \bibnamefont {Ardekani}},\ }\bibfield  {title} {\bibinfo {title} {Motion of a model swimmer near a weakly deforming interface},\ }\href@noop {} {\bibfield  {journal} {\bibinfo  {journal} {Journal of Fluid Mechanics}\ }\textbf {\bibinfo {volume} {824}},\ \bibinfo {pages} {42} (\bibinfo {year} {2017})}\BibitemShut {NoStop}%
\bibitem [{\citenamefont {Mathijssen}\ \emph {et~al.}(2016{\natexlab{b}})\citenamefont {Mathijssen}, \citenamefont {Doostmohammadi}, \citenamefont {Yeomans},\ and\ \citenamefont {Shendruk}}]{mathijssen2016hydrodynamics}%
  \BibitemOpen
  \bibfield  {author} {\bibinfo {author} {\bibfnamefont {A.~J.}\ \bibnamefont {Mathijssen}}, \bibinfo {author} {\bibfnamefont {A.}~\bibnamefont {Doostmohammadi}}, \bibinfo {author} {\bibfnamefont {J.~M.}\ \bibnamefont {Yeomans}},\ and\ \bibinfo {author} {\bibfnamefont {T.~N.}\ \bibnamefont {Shendruk}},\ }\bibfield  {title} {\bibinfo {title} {Hydrodynamics of micro-swimmers in films},\ }\href@noop {} {\bibfield  {journal} {\bibinfo  {journal} {Journal of Fluid Mechanics}\ }\textbf {\bibinfo {volume} {806}},\ \bibinfo {pages} {35} (\bibinfo {year} {2016}{\natexlab{b}})}\BibitemShut {NoStop}%
\bibitem [{\citenamefont {Spagnolie}\ \emph {et~al.}(2015)\citenamefont {Spagnolie}, \citenamefont {Moreno-Flores}, \citenamefont {Bartolo},\ and\ \citenamefont {Lauga}}]{spagnolie2015geometric}%
  \BibitemOpen
  \bibfield  {author} {\bibinfo {author} {\bibfnamefont {S.~E.}\ \bibnamefont {Spagnolie}}, \bibinfo {author} {\bibfnamefont {G.~R.}\ \bibnamefont {Moreno-Flores}}, \bibinfo {author} {\bibfnamefont {D.}~\bibnamefont {Bartolo}},\ and\ \bibinfo {author} {\bibfnamefont {E.}~\bibnamefont {Lauga}},\ }\bibfield  {title} {\bibinfo {title} {Geometric capture and escape of a microswimmer colliding with an obstacle},\ }\href@noop {} {\bibfield  {journal} {\bibinfo  {journal} {Soft Matter}\ }\textbf {\bibinfo {volume} {11}},\ \bibinfo {pages} {3396} (\bibinfo {year} {2015})}\BibitemShut {NoStop}%
\bibitem [{\citenamefont {Desai}\ \emph {et~al.}(2018)\citenamefont {Desai}, \citenamefont {Shaik},\ and\ \citenamefont {Ardekani}}]{desai2018hydrodynamics}%
  \BibitemOpen
  \bibfield  {author} {\bibinfo {author} {\bibfnamefont {N.}~\bibnamefont {Desai}}, \bibinfo {author} {\bibfnamefont {V.~A.}\ \bibnamefont {Shaik}},\ and\ \bibinfo {author} {\bibfnamefont {A.~M.}\ \bibnamefont {Ardekani}},\ }\bibfield  {title} {\bibinfo {title} {Hydrodynamics-mediated trapping of micro-swimmers near drops},\ }\href@noop {} {\bibfield  {journal} {\bibinfo  {journal} {Soft Matter}\ }\textbf {\bibinfo {volume} {14}},\ \bibinfo {pages} {264} (\bibinfo {year} {2018})}\BibitemShut {NoStop}%
\bibitem [{\citenamefont {Blake}(1971)}]{blake1971note}%
  \BibitemOpen
  \bibfield  {author} {\bibinfo {author} {\bibfnamefont {J.~R.}\ \bibnamefont {Blake}},\ }\bibfield  {title} {\bibinfo {title} {A note on the image system for a stokeslet in a no-slip boundary},\ }\href@noop {} {\bibfield  {journal} {\bibinfo  {journal} {Mathematical Proceedings of the Cambridge Philosophical Society}\ }\textbf {\bibinfo {volume} {70}},\ \bibinfo {pages} {303} (\bibinfo {year} {1971})}\BibitemShut {NoStop}%
\bibitem [{\citenamefont {Abramowitz}\ and\ \citenamefont {Stegun}(1965)}]{abramowitz1965handbook}%
  \BibitemOpen
  \bibfield  {author} {\bibinfo {author} {\bibfnamefont {M.}~\bibnamefont {Abramowitz}}\ and\ \bibinfo {author} {\bibfnamefont {I.~A.}\ \bibnamefont {Stegun}},\ }\href@noop {} {\emph {\bibinfo {title} {Handbook of mathematical functions: with formulas, graphs, and mathematical tables}}}\ (\bibinfo  {publisher} {Dover Publications},\ \bibinfo {address} {New York},\ \bibinfo {year} {1965})\BibitemShut {NoStop}%
\bibitem [{\citenamefont {Oldham}\ \emph {et~al.}(2009)\citenamefont {Oldham}, \citenamefont {Myland},\ and\ \citenamefont {Spanier}}]{oldham2009atlas}%
  \BibitemOpen
  \bibfield  {author} {\bibinfo {author} {\bibfnamefont {K.~B.}\ \bibnamefont {Oldham}}, \bibinfo {author} {\bibfnamefont {J.}~\bibnamefont {Myland}},\ and\ \bibinfo {author} {\bibfnamefont {J.}~\bibnamefont {Spanier}},\ }\href@noop {} {\emph {\bibinfo {title} {An atlas of functions: with equator, the atlas function calculator}}}\ (\bibinfo  {publisher} {Springer},\ \bibinfo {address} {New York},\ \bibinfo {year} {2009})\BibitemShut {NoStop}%
\bibitem [{\citenamefont {Olver}\ \emph {et~al.}(2010)\citenamefont {Olver}, \citenamefont {Lozier}, \citenamefont {Boisvert},\ and\ \citenamefont {Clark}}]{olver2010nist}%
  \BibitemOpen
  \bibfield  {author} {\bibinfo {author} {\bibfnamefont {F.~W.~J.}\ \bibnamefont {Olver}}, \bibinfo {author} {\bibfnamefont {D.~W.}\ \bibnamefont {Lozier}}, \bibinfo {author} {\bibfnamefont {R.~F.}\ \bibnamefont {Boisvert}},\ and\ \bibinfo {author} {\bibfnamefont {C.~W.}\ \bibnamefont {Clark}},\ }\href@noop {} {\emph {\bibinfo {title} {{NIST} Handbook of Mathematical Functions}}}\ (\bibinfo  {publisher} {Cambridge University Press},\ \bibinfo {address} {Cambridge},\ \bibinfo {year} {2010})\BibitemShut {NoStop}%
\end{thebibliography}
\end{document}